\documentclass[a4paper,12pt]{article}
\usepackage{graphicx,amssymb,bm,latexsym,color,epsf,cite,ulem,verbatim, physics, mathrsfs,lipsum, amsfonts,amsthm,stmaryrd, bm, hhline, float, empheq, tcolorbox, hyperref, etoolbox, threeparttable, enumitem,upgreek, subcaption, fancyhdr,cases, bigints, wrapfig}
\tcbuselibrary{theorems}
\DeclareMathAlphabet\mathbfcal{OMS}{cmsy}{b}{n}
\newcommand{\ffrac}[2]{\ensuremath{\frac{\displaystyle #1}{\displaystyle #2}}}

\usepackage[title]{appendix}

 \usepackage{caption, setspace}
 \usepackage{lineno}

\makeatletter
\@addtoreset{equation}{section}

\makeatother
\newcommand{\vs}[1]{\vspace{#1 mm}}
\newcommand*\di{\mathop{}\!\mathrm{d}}

{
\addtolength{\skip\footins}{1pc plus 1pt}
\setlength{\footnotesep}{1pc}

\usepackage[margin=1in,includehead,includefoot,headheight=0.5in,headsep=0.2in,footskip=0.47in,heightrounded]{geometry}

\pagestyle{plain}
\textwidth 160mm
\textheight 240mm
\topmargin -22mm
\oddsidemargin 0mm
\begin{document}
	\newgeometry{top = 2.5cm,bottom = 2.5cm, left=2cm, right=2cm}
\pagestyle{fancy}
\fancyhf{}
\fancyhead[R]{MITP-22-029}
\fancyfoot[C]{ }
\renewcommand{\headrulewidth}{0pt}
\cfoot{}

{  }
\clearpage

	\begin{center}
	
		{\LARGE  ${}$\vs{+3.5}On the possibility of a novel (A)dS/CFT relationship \vs{+1.8}\\emerging in  Asymptotic Safety}                                     		\vs{10}
		
		{\large
			Renata Ferrero\footnote{e-mail address: rferrero@uni-mainz.de}
			and Martin Reuter\footnote{e-mail address: reutma00@uni-mainz.de}$^{}$
		} \\
		\vs{10}
	{\textit{Institute of Physics (THEP), University of Mainz,
				\\Staudingerweg 7, D-55128 Mainz, Germany}
		}
	
	\end{center}
\vs{5}

\setcounter{footnote}{0} 

\begin{abstract}
Quantum Einstein Gravity (QEG), nonperturbatively renormalized by means of a certain asymptotically safe renormalization group (RG) trajectory, is explored by solving its scale dependent effective field equations and  embedding the family of emerging 4-dimensional spacetimes into a single 5-dimensional manifold, which thus encodes the complete information about all scales. By construction the latter manifold is furnished with a natural foliation. Heuristically, its leaves are interpreted as physical spacetime  observed on different scales of the experimental resolution.
Generalizing earlier work on the embedding of $d$-dimensional Euclidean QEG spacetimes in ($d+1$)-dimensional flat or Ricci flat manifolds,  we admit Lorentzian signature in this paper and we consider  embeddings in arbitrary ($d+1$)-dimensional Einstein spaces. Special attention is paid to the sector of maximally symmetric metrics, and the fundamental definition of QEG in $d = 4$ that employs the cross-over trajectory connecting the non-Gaussian to the Gaussian RG fixed point. Concerning the embedding of the resulting family of 4D de Sitter solutions with a running Hubble parameter, we find that there are only two possible 5D spacetimes, namely the anti-de Sitter manifold AdS$_5$ and the de Sitter manifold dS$_5$. To arrive at this result essential use is made of the monotone scale dependence of the running cosmological constant featured by the gravitational effective average action. We show that if the scale invariance of the QEG fixed points extends to full conformal invariance,  the 5D picture of the resulting  geometric and field theoretic structure displays a novel kind of ``AdS/CFT correspondence''. While strongly reminiscent of the usual string theory-based AdS/CFT correspondence, also clear differences are found.
\end{abstract}
\thispagestyle{fancy}

\newpage
\restoregeometry
\setcounter{page}{1}

\pagestyle{fancy}
\fancyhf{}
\fancyhead[R]{}
\renewcommand{\headrulewidth}{0pt}
\cfoot{\thepage}
\pagestyle{plain}

\section{Introduction}
The gravitational Effective Average Action (GEAA) is a versatile framework of quantum field theory for the Background Independent and generally covariant quantization of gravity and matter fields  coupled to it \cite{Martin}. The concepts involved in, and practical tools provided by this approach are fully nonperturbative and do not assume a pre-existing spacetime. Being rooted in the functional renormalization group, the GEAA describes gravitational systems in terms of a one-parameter family of effective field theories. They describe the properties of a dynamically generated spacetime, and the dynamics of gravitational and matter fluctuations therein on different resolution scales.

While the bare field representing gravity at the microscopic level is not restricted to be a metric tensor \cite{Jan1, Jan2, Harst0, Harst1, Harst2}, the expectation values encoded in the GEAA functionals include a scale dependent spacetime metric \cite{Carlo1, Pagani}. Its dependence on the coarse graining scale gives rise to a fractal-like picture of the quantum gravity spacetimes at the mean field level \cite{Lauscher}. Thereby the emergence of a classic world from the quantum regime hinges on whether or not the renormalization group (RG) evolution comes to a halt eventually \cite{Frank}.

\bigskip

\noindent\textbf{(1)} Separately for each RG scale $k \in \mathbb{R}^+$, the respective GEAA functional $\Gamma_k$ implies a  quantum corrected variant of Einstein's equation;  its solutions are the resolution dependent metrics $g_{\mu \nu}^k$. They are different for different scales usually, but establish (pseudo-) Riemannian structures on one and the same smooth manifold, $\mathscr{M}_4$.

If we regard the primary RG trajectory on the theory space under consideration, $\mathscr{T}$, as a map $\mathbb{R}^+ \to \mathscr{T}$, $k \mapsto\Gamma_k$, then the 	``running'' solution to the scale dependent Einstein equations can be seen as an associated map from $\mathbb{R}^+$ into the space of metrics on $\mathscr{M}_4$. Thereby the association $k\mapsto g_{\mu \nu}^k$ describes the family of (pseudo-) Riemannian structures $\Big(\mathscr{M}_4,\, g_{\mu \nu}^k\Big)$ which quantum spacetime displays for different values of the RG parameter. Heuristically, we may think of $g_{\mu \nu}^k$ as the effective metric which is detected in experiments that involve a typical momentum scale of the order of $k$.
\bigskip

\noindent\textbf{(2)} In  ref.  \cite{Renata1}, henceforth referred to as [I], we proposed a new way of representing and analyzing the family of metrics $g_{\mu \nu}^k$ that furnish the same, given 4-dimensional manifold $\mathscr{M}_4$. The idea is to interpret the 4D spacetimes $\Big(\mathscr{M}_4,\, g_{\mu \nu}^k\Big)$, $k \in \mathbb{R}^+$, as different slices through a single 5-dimensional (pseudo-) Riemannian manifold: $\Big(\mathscr{M}_5,\, {}^{(5)}g_{IJ}\Big)$. Thereby the $g_{\mu \nu}^k$'s are related to the 5D metric  $ {}^{(5)}g_{IJ}$ by an \textit{isometric embedding} of the 4D slices into $\mathscr{M}_5$. Stated the other way around, $\Big(\mathscr{M}_5,\, {}^{(5)}g_{IJ}\Big)$ is a single \textit{foliated manifold}, the leaves of whose foliation describe the spacetime at different values of the RG parameter.

As $k$ is an inverse coarse graining scale on $\mathscr{M}_4$, the manifold $\mathscr{M}_5$ has the interpretation of a ``scale-space-time'' similar in spirit to the ones considered in \cite{Nottale}. In addition to the usual coordinates of an event, $x^\mu$, coordinates on $\mathscr{M}_5$ include a value of the RG parameter $k$, or an appropriate function thereof.

In a special system of coordinates that is adapted to the foliation, points on $\mathscr{M}_5$ have coordinates $x^I=(k, x^\mu)$, and the metrics $g_{\mu \nu}(k, x^\rho) \equiv g_{\mu \nu}^k (x^\rho)$ can be identified directly with 10 out of the 15 independent components which $^{(5)}g_{IJ}$ possesses.

Now, the intriguing question is about the additional 5 components of $^{(5)}g_{IJ}$ that are \textit{not} provided by the 4D flow. In ref. [I] we discussed the possibility that  there could exist mathematically or physically distinguished ways of fixing these additional components, the idea being that the (then unique) 5D geometry encapsulates not only the entirety of the 4D geometries, but enriches it by additional physics contents.

Within the restricted setting of ref. [I], we were indeed able to identify an additional piece of physics information which gets encoded by the very existence of $\Big(\mathscr{M}_5,\, {}^{(5)}g_{IJ}\Big)$ of a certain type, namely the property that the running cosmological constant $\Lambda(k)$ is a monotonically decreasing function of $k$.\footnote{This principle is similar to, but not identical with the GEAA-based proposal for a generally applicable concept of a $C$-function in ref.\cite{Becker}.}

Assuming this to be the case, we showed that 4D \textit{Euclidean} spacetimes can always be embedded in a $\mathscr{M}_5$ which is Ricci flat, or even Riemann flat if the 4D spacetimes are maximally symmetric.

\bigskip

\noindent\textbf{(3)} In the present work, we are going to extend these investigations in two directions: First, we allow the higher dimensional manifold $\mathscr{M}_5$ to be an arbitrary \textit{Einstein space}, and second, we admit the possibility that the spacetimes to be embedded, $\Big(\mathscr{M}_4,\, g_{\mu \nu}^k\Big)$, have a \textit{Lorentzian signature}, a prime example being a stack of de Sitter spaces dS$_4$ with a $k$-dependent Hubble parameter.

The two extensions are not unrelated. Namely, aiming at a \textit{global} embedding of the dS$_4$'s into a 5-dimensional manifold, one has to face general theorems \cite{nash1, kuiper, Friedman1, Nash} which tell us that the corresponding $\Big(\mathscr{M}_5,\, {}^{(5)}g_{IJ}\Big)$ cannot be flat \cite{Kasner, Rosen} or Ricci flat \cite{Campbell, Magaard}.

Thus, being particularly interested in the physically relevant case of 4D Lorentzian spacetimes, in the present work we examine the most natural generalization beyond flat and Ricci flat $\mathscr{M}_5$'s, namely 5D Einstein spaces.

\bigskip

\noindent\textbf{(4)} Like in [I], also in this paper we shall illustrate the general considerations by examples taken from Asymptotic Safety, more precisely from Quantum Einstein Gravity (QEG), the nonperturbatively renormalized quantum theory whose basic degrees of freedom are carried by a bare metric field \cite{Weinberg, Martin, Oliver, Frank1, Oliver2, Oliver3}.

\begin{figure}[t]
	\centering
	\includegraphics[scale=0.225]{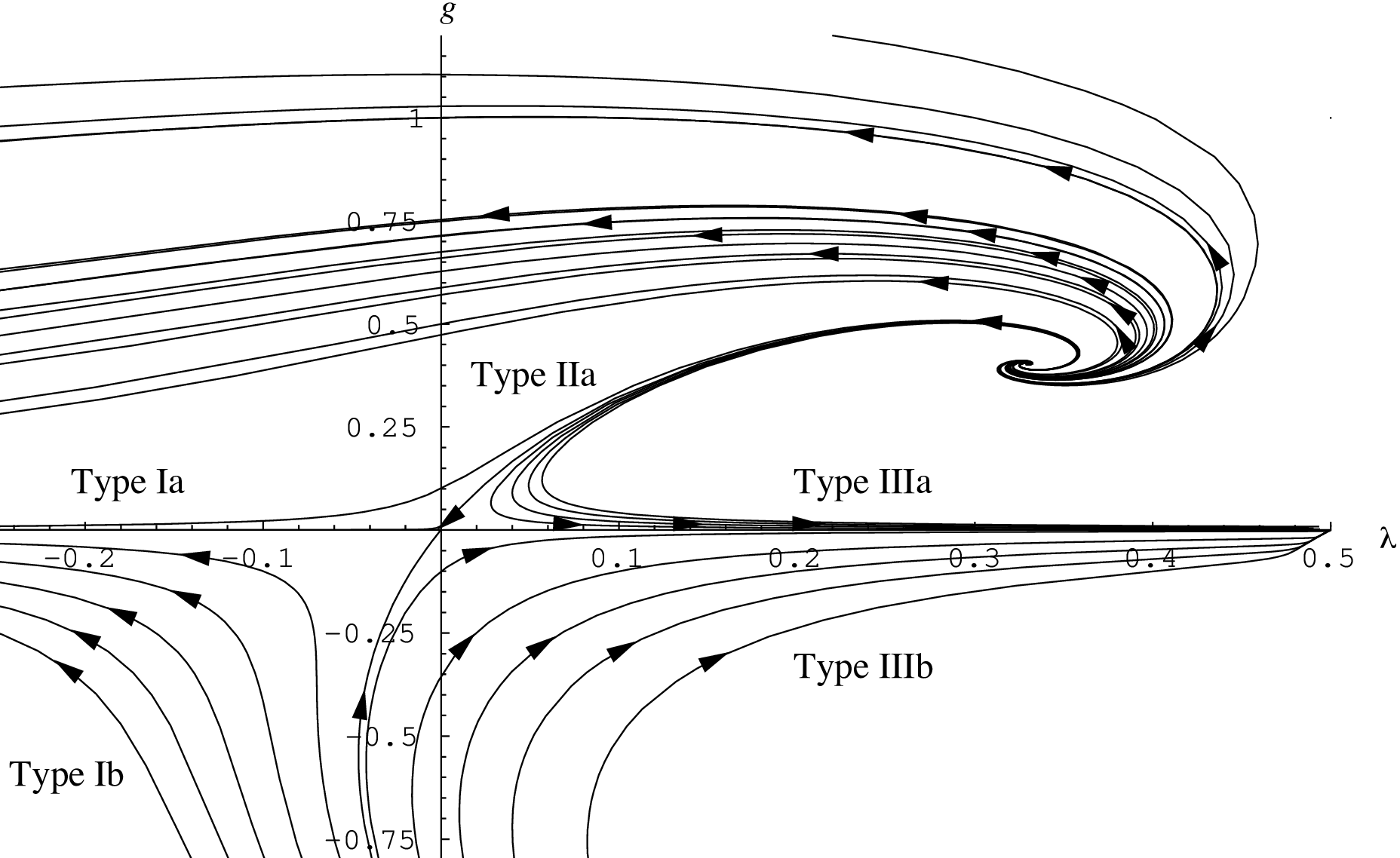}
	\caption{The RG flow on the theory space of the Einstein-Hilbert truncation. The arrows point in the direction of decreasing $k$-values. The phase portrait is dominated by the NGFP in the first quadrant and the GFP at the origin. (Taken from \cite{Frank1}.)\vspace{-0.5cm}}\label{fig:EH-flow}
\end{figure}

To keep the presentation as simple as possible, all explicit examples are based upon the prototypical RG flow obtained from the 4D Einstein-Hilbert truncation \cite{Martin,Oliver,Frank1}:
\begin{equation}
	\Gamma_k = \Big(16\pi G(k)\Big)^{-1}\bigintsss \di ^4x \sqrt{g}\Big(-R+2\Lambda(k)\Big) + \cdots\;.
	\label{trunc}
	\end{equation}
In Figure \ref{fig:EH-flow} we display the  phase portrait which is obtained after inserting the ansatz into the functional flow equation.\footnote{See refs.\cite{Percacci, Frank} for a comprehensive account of the corresponding calculational techniques. We also refer to \cite{Wetterich-2, Wetterich-1, Wetterich-0, Wetterich+1, Wetterich+2} for the detailed constructions that led to the effective average action on flat space. Furthermore, a brief introduction to the main aspects of the GEAA can be found in the Appendix.} On the 2D theory space,  the dimensionless cosmological constant $\lambda(k) = \Lambda (k) /k^2$ and Newton constant $g(k) = k^2 G(k)$ serve as coordinates.

The RG flow is dominated by the non-Gaussian fixed point (NGFP) at $(g_\ast, \lambda_\ast) \neq 0$, and the (so-called) Gaussian fixed point (GFP) with $g_\ast = 0 = \lambda_\ast$.  In the present paper we shall focus on two classes of RG trajectories, namely those of Type IIIa and IIa, respectively. The former have a strictly positive cosmological constant $\Lambda(k)$ on all scales, and this includes the physical value  $k = 0$ at the trajectory's end point, $\Lambda(k = 0) \equiv \Lambda_0 >0$. The Type IIa comprises a single trajectory only, the separatrix seen in the phase portrait of Figure \ref{fig:EH-flow}. It describes the cross-over from the NGFP to the GFP, and gives rise to a vanishing renormalized cosmological constant: $\Lambda(0) \equiv \Lambda_0 = 0$.

The RG flow of Figure \ref{fig:EH-flow} applies not only to pure QEG, but also to QEG coupled to  a large variety of possible matter systems  \cite{Dou, Percacci, Frank}.
Furthermore, this flow is valid both in the Euclidean and the Lorentzian setting, see refs.\cite{renata-spec, Platania, Knorr} for a recent discussion of this point.

As for the very existence of the NGFP, more than two decades of work on QEG within a  considerable number of truncations of ever increasing generality and complexity led to the basically   inescapable conclusion that QEG does indeed possess the fixed point which is necessary for its nonperturbative renormalizability \cite{Martin, Oliver,  Frank1, Oliver2, Oliver3, 10people}.

And yet, some of the more specific properties of the QEG fixed points still remain to be established. For instance, it is  an open question whether their scale invariance  extends to a full-fledged conformal invariance. In the special case of $d = 2$ spacetime dimensions, however, this question has been already answered in the affirmative, and moreover a unitary 2D conformal field theory (CFT) has been identified which governs the fixed point theory  \cite{Andi}.

\bigskip
\noindent\textbf{(5)} One of the motivations for the research program initiated in [I] was our conjecture that there should be rather a close relationship between Asymptotic Safety on the one side, and the string theory-based AdS/CFT correspondence  on the other \cite{Maldacena, Gubser, Witten, Erd}.

Given the scope of the GEAA approach and its applicability to arbitrary systems of fields, it is clear that it  also addresses the questions about the conformal field theory (CFT) on the AdS$_5$ boundary which the AdS/CFT framework  answers by holographic means. Therefore a presumably harder, but in principle exact GEAA-based calculation should be able to come up with specific answers to the same questions, and clearly it would be extremely interesting to see whether the respective answers agree.

Closely related is the following question concerning the relationship between the Asymptotic Safety and the AdS/CFT frameworks in general: Is it possible to ``derive'' a certain kind of, possibly non-standard AdS/CFT correspondence by applying the GEAA approach to a specific system of 4D gravity + matter fields,  computing its running actions $\Gamma_k$ and background spacetimes $g_{\mu \nu}^k$ by solving the  functional RG and effective Einstein equations, and then embedding the 4D metrics into  a  5-dimensional one?

It is an intriguing possibility that by following these steps of the geometrization program one might be led to a specific solution of the general GEAA flow and field equations which describes a 5D   setting with a bulk/surface relationship similar to that of the well-known AdS/CFT conjecture.

In this paper we present a first indication which indeed points in precisely this direction. For the time being, the results are mostly restricted to the geometric aspects of the correspondence however.

\bigskip
\noindent\textbf{(6) Plan of this paper.}  In Section \ref{sec:2} we are going to investigate the possibility of embedding scale histories of $d$-dimensional effective metrics $g_{\mu \nu}^k$ in a unique ($d+1$)-dimensional manifold. In fact, most discussions in this paper are valid for arbitrary dimensionalities $d$. Working within the same class of higher-dimensional metrics $^{(d+1)}g_{IJ}$ as in the preceding paper [I], we now investigate the particularly relevant case where $^{(d+1)}g_{IJ}$ is Einstein, and both $g_{\mu \nu}^k$ and $^{(d+1)}g_{IJ}$ are Lorentzian.

In Section \ref{sec:3}, we  impose maximum symmetry on $g_{\mu \nu}^k$, and find that under this condition $^{(d+1)}g_{IJ}$ can be chosen maximally symmetric, too. This leaves us with two potential candidates for an embedding spacetime, namely $\mathscr{M}_{d+1}=$ AdS$_{d+1}$ and  $\mathscr{M}_{d+1}=$ dS$_{d+1}$, respectively.

In Section \ref{sec:4}, we then prove that both options are viable actually, i.e., that there exists an admissible coordinate transformation that relates the RG parameter $k$ to the coordinate which labels the leaves of the foliation displayed by $^{(d+1)}g_{IJ}$. In this step, essential use will be made of the monotone $k$-dependence of the running cosmological constant.

In Section \ref{sec:5}, we analyze the global properties of the embeddings obtained, and we show that the picture of a certain ``(A)dS/CFT correspondence''  emerges from the RG flow of QEG, thanks to its Asymptotic Safety.

In Section \ref{sec:new6} we compare this non-standard correspondence to the usual one based upon string theory, highlighting their similarities and main differences.

Finally, Section \ref{sec:6} contains a brief summary and our conclusions.

Furthermore, we provide a brief introduction to the GEAA in the Appendix, as well as additional information on the RG trajectories employed in this paper.
\vspace{-0.4cm}
\section{Embedding in Einstein manifolds}\label{sec:2}
To place the metric of the embedding manifold $\Big(\mathscr{M}_{d+1},\, {}^{(d+1)}g_{IJ}\Big)$ in a broader context, we start from $(d+1)$-dimensional line elements of the same form as in our investigation of the flat and Ricci flat embeddings in [I], namely:
\begin{equation}
	\boxed{	{}^{(d+1)}g_{IJ}(x^K) \di x^I \di x^J = \Omega ^2 (\gamma)\; \Big[\varepsilon \,(\di \gamma)^2 + g^\text{R}_{\mu \nu}(x^\rho) \di x^\mu \di x^\nu\Big]\;.}
	\label{d+1}
\end{equation}
Here $\Omega(\gamma)$ is an arbitrary  conformal factor, and  coordinates $x^K \equiv \left(x^0 = \gamma, \;x^\mu\right)$ are used.\footnote{Our conventions are the same as in [I]. In particular $\mu, \nu, \cdots \in \{1,2,\cdots,d\}$ and $I,J,\cdots \in \{0,1,2,\cdots,d\}$. Thereby $x^0$ is the scale-, but not necessarily a time-coordinate.} The ordinary event-coordinates are denoted $x^\mu$, while $\gamma$ is the additional scale coordinate. 
In [I] we were led to the ansatz \eqref{d+1} by starting out from a generic metric in ``scale-ADM'' form and then imposing the restrictions of a vanishing shift vector and a $x^\mu$-independent lapse function.

Later on we shall try to relate $\gamma$ to the RG parameter $k$ by a diffeomorphic relationship $\gamma = \gamma(k)$ in such a way that $\Omega^2(\gamma(k)) g_{\mu \nu}^\text{R} (x^\rho)$ coincides with the 	``running metrics'' $g_{\mu \nu}^k(x^\rho)$ obtained by means of the GEAA methods.

The sign factor $\varepsilon= \pm 1$ allows the scale variable $\gamma$ to be introduced either as a time or a space coordinate. For now, this choice is unrelated to the signature of the $d$-dimensional 	``reference  metric'', $g^\text{R}_{\mu \nu}(x^\rho)$. In fact, this signature is left upon at this point, and $g^\text{R}_{\mu \nu}(x^\rho)$ can be a Lorentzian or a Euclidean metric.

In either case, the Ricci tensor ${}^{(d+1)}R^I{}_J$ of the above metric $	{}^{(d+1)}g_{IJ}$ has the following components:
\begin{subequations}
	\begin{align}
		{}^{(d+1)}R^0{}_0 & = -\varepsilon \;d \;\Omega^{-2}\left[\frac{\ddot{\Omega}}{\Omega}- \left(\frac{\dot \Omega}{\Omega}\right)^2\right]\;,\label{r00} \\
		{}^{(d+1)}R^0{}_\mu & = 0, \qquad 	{}^{(d+1)}R^\mu{}_0  = 0\;,\\
		{}^{(d+1)}R^\mu{}_\nu & = \Omega^{-2}\left\{ R^\mu{}_\nu - \varepsilon\; \delta^\mu{}_\nu\left[\frac{\ddot{\Omega}}{\Omega}+(d-2) \left(\frac{\dot \Omega}{\Omega}\right)^2\right] \right\} \;. \label{rmix}
	\end{align}
\end{subequations}
Here and in the following dots indicate derivatives with respect to $\gamma$, and $R^\mu{}_\nu$ denotes the $d$-dimensional Ricci tensor\footnote{In conformity with the conventions adopted in [I], we denote higher-dimensional geometric quantities (such as the curvature scalar ${}^{(d+1)}R$, say) by a prepended label $(d+1)$, while all objects without this label are $d$-dimensional ones referring to $\mathscr{M}_d$. For example, $R_{\mu \nu}$ denotes the Ricci tensor built from $g_{\mu \nu}$ in $d$ dimensions, whereas ${}^{(d+1)}R_{\mu \nu}$ are the $\mu$-$\nu$-components of the tensor ${}^{(d+1)}R_{IJ}$ which derives from ${}^{(d+1)}g_{I J}$.} belonging to $g^\text{R}_{\mu \nu}(x^\rho)$.

\bigskip
\noindent \textbf{(1) The reference metric.} Concerning the family of metrics $\left\{g_{\mu \nu}^k(x^\rho) \,|\,k \in \mathbb{R}^+\right\}$ delivered by the GEAA, we assume that they are related to a  trajectory of Type IIIa or IIa of the Einstein-Hilbert truncation. Since the latter entails the classical-looking effective field equation $G_{\mu \nu}[g_{\alpha \beta}^k] = -\Lambda(k)g_{\mu \nu}^k$, its solutions $g_{\mu \nu}^k$ are $d$-dimensional Einstein metrics. As $g_{\mu \nu}^k$ is related to $g^\text{R}_{\mu \nu}$ by the $x^\rho$-independent conformal factor $\Omega^2$, we are led to make the following assumptions about the $k$-independent reference metric:
\bigskip

\textbf{(i)} \textit{The $d$-dimensional reference metric $g^\text{R}_{\mu \nu}(x^\rho)$ is Einstein}:
\begin{equation}
	R^{\mu}{}_\nu \left[g^\text{R}_{\alpha \beta}\right] = \frac{2}{(d-2)}\;\Lambda_\text{R}\;\delta^\mu{}_\nu
\end{equation}

\textbf{(ii)} \textit{The corresponding cosmological constant is strictly positive}:
\begin{equation}
	\Lambda_\text{R} >0\;.
\end{equation}
Often it will be convenient in the following to trade $\Lambda_\text{R}$ for the parameter
\begin{equation}
	H_\text{R} \equiv \sqrt{\frac{2\; \Lambda_\text{R}}{(d-1)(d-2)}}\;,
\end{equation}
in terms of which $\Lambda_\text{R} \equiv \frac{1}{2} (d-1)(d-2)H_\text{R}^2$, and 
\begin{equation}
	R^{\mu}{}_\nu \left[g^\text{R}_{\alpha \beta}\right] = (d-1)\;H_\text{R}^2\;\delta^\mu{}_\nu\;,
\end{equation}
\begin{equation}
	R \left[g^\text{R}_{\alpha \beta}\right] =	R^{\mu}{}_\mu \left[g^\text{R}_{\alpha \beta}\right] = d\;(d-1)\;H_\text{R}^2\;.
\end{equation}
For the special case of the de Sitter solution, the parameter $H_\text{R}$ happens to coincide with its Hubble constant. However, for the time being we consider  fully generic Einstein metrics $g_{\mu \nu}^\text{R}$.

\bigskip
\noindent \textbf{(2) The Einstein condition.} Let us now impose the condition that the embedding metric in $d+1$ dimensions, eq.\eqref{d+1}, too, is an Einstein metric:
\begin{equation}
	\boxed{{}^{(d+1)}R^{I}{}_J = C\; \delta^I {}_J\;.}
\end{equation}
Here $C$  is essentially the higher-dimensional cosmological constant: \mbox{$C = \left(\frac{2}{d-1}\right)\mathbf{{}^{(d+1)}\Lambda}$}. The Einstein condition leads to the following constraints on $\varepsilon$ and $\Omega (\gamma)$:
\begin{subequations}
	\begin{align}
		\varepsilon\; d\; \dot \Omega^2 - \varepsilon\; d\; \Omega \; \ddot \Omega\; &=\; C\; \Omega^4\;,\\
		(d-1)\; H_\text{R}^2 \; \Omega^2 - \varepsilon\;(d-2) \; \dot \Omega^2 -\varepsilon \; \Omega\; \ddot{\Omega}\;& = \;C\; \Omega^4\;.
	\end{align}
\end{subequations}
To proceed, it is advantageous to replace these  differential equations by their sum and difference, respectively, and to express them in terms of the function
\begin{equation}
	\omega (\gamma) \equiv 1/\Omega(\gamma)\;.
\end{equation}
We obtain, respectively,
\begin{subequations}
	\begin{align}
		(d+1)\; \dot \omega \;\ddot{\omega} -2\; d\; \dot{\omega}^2+ \varepsilon\; (d-1) \;H_\text{R}^2\; \omega^2\; &=\;2\; \varepsilon\; C\;,\label{23a}\\
		\ddot{\omega} -\varepsilon\; H_\text{R}^2 \; \omega\;&=\;0\;. \label{23b}
	\end{align}
\end{subequations}

The second equation above is easily solved. Depending on whether $\gamma$ is a space or a time coordinate, we obtain, with integration constants $\alpha_1$ and $\alpha_2$:
\begin{empheq}[box=\fbox]{align}
	\varepsilon=+1&: \qquad \omega(\gamma) = \alpha_1\; \sinh(H_\text{R}\,\gamma) +\alpha_2\; \cosh(H_\text{R}\, \gamma)\;,\label{24}\\
	\varepsilon=-1&: \qquad \omega(\gamma) = \alpha_1\; \sin(H_\text{R}\,\gamma) +\alpha_2\; \cos(H_\text{R}\, \gamma)\;.
\end{empheq}

For practical calculations it is often better not to use the explicit solutions but simply to exploit that \eqref{23b} admits the first integral
\begin{equation}
	\dot \omega^2 - \varepsilon\; H_\text{R}^2\; \omega^2 = \text{const} \equiv E\;.
\end{equation}
If desired, one can express the $\gamma$-independent ``energy'' $E$ in terms of the integration constants $\alpha_1$ and $\alpha_2$ according to 
\begin{equation}
	E = \Big(\alpha_1^2-\varepsilon\; \alpha_2^2\Big)\; H_\text{R}^2\;.
\end{equation}

Therefore, turning to the first differential equation \eqref{23a} now, we are entitled to make the following substitutions there:
\begin{equation}
	\dot \omega^2 = E + \varepsilon\; H_\text{R}^2 \;\omega^2 \qquad \text{and} \qquad \omega\; \ddot{\omega} = \varepsilon\; H_\text{R}^2 \;\omega^2\;.
\end{equation}
As a result, the dependence on $\omega (\gamma)$ drops out completely from \eqref{23a}. What remains is a condition that relates $C$ to the constants of integration:
\begin{eqnarray}
	C\;&=&\;-\varepsilon\; d \; E\\
	&=&\;d\; \Big(-\varepsilon\; \alpha_1^2+\alpha_2^2\Big)\; H_\text{R}^2
\end{eqnarray}
In terms of the conventionally normalized cosmological constant in $(d+1)$ dimensions, $\mathbf{	{}^{(d+1)}\Lambda}$,  this value of $C$ amounts to 
\begin{eqnarray}
	\mathbf{	{}^{(d+1)}\Lambda}\;&=&\;\frac{1}{2}\;(d-1)\;C\\
	&=&\;\frac{1}{2}\;d\;(d-1)\;\; \Big[-\varepsilon\; \alpha_1^2+\alpha_2^2\Big]\; H_\text{R}^2\;.\label{33}
\end{eqnarray}
Obviously, $\mathbf{	{}^{(d+1)}\Lambda}$ can have either sign, depending on the relative magnitude of $\alpha_1$ and $\alpha_2$, and on the factor $\varepsilon$.

The case of a vanishing $\mathbf{	{}^{(d+1)}\Lambda} = 0$ is included here as well, and it leads us back to the Ricci flat metrics considered in [I]. By eq.\eqref{33}, this case is seen to require that $\varepsilon=+1$ and $\alpha_1 = \pm \alpha_2$. Hence, from \eqref{24}, we obtain $\omega(\gamma) \propto \exp\left(\pm H_\text{R}\gamma\right)$, and choosing the upper sign this yields  the metric given in eq.(5.21) of [I] for the presently assumed positive sign of $\Lambda_\text{R}$, $\sigma = +1$.

However, while the discussion in [I] assumed \textit{Euclidean} metrics $g^k_{\mu \nu}$, we now see that under the same conditions it is  possible to embed also \textit{Lorentzian} spacetimes $\Big(\mathscr{M}_d, \,g^k_{\mu \nu}\Big)$ in a Ricci flat $\mathscr{M}_{d+1}$.

The point to be noted here is that the above investigation of  differential equations is of a local nature and hence yields criteria for the existence of \textit{local} embeddings only. But, importantly, in trying to extend a local embedding to a \textit{global} one, the signature of $\Big(\mathscr{M}_d, \,g^k_{\mu \nu}\Big)$ is of crucial importance.

In this sense, the spacetimes $\mathscr{M}_{d+1}$ found above merely have  the status of \textit{candidates} for a global embedding.

\bigskip
\noindent\textbf{(3) The Riemann tensor $\bf{{}^{(d+1)}R^{IJ}{}_{KL}}$.} Before turning to specific solutions, let us look at the Riemannian tensor ${}^{(d+1)}R^{IJ}{}_{KL}$ of the higher-dimensional metrics  ${}^{(d+1)}g_{IJ}$ based upon an arbitrary $d$-dimensional Einstein metric $g^\text{R}_{\mu \nu}$. For metrics of the type \eqref{d+1}, its only nonzero components, up to the usual symmetries, are
\begin{subequations}
	\begin{align}
		{}^{(d+1)}R^{0\mu}{}_{0\nu} &=C\;d^{-1} \; \delta^\mu_{\nu}\;,\label{40a}\\
		{}^{(d+1)}R^{\mu \nu}{}_{\rho \sigma} &= C\; d^{-1}\; \Big[\delta^\mu_\rho\; \delta^\nu_\sigma - \delta_\sigma ^\mu\; \delta _\rho ^\nu\Big] + \bigg\{R^{\mu \nu}{}_{\rho \sigma} - H_\text{R}^2\Big[\delta^\mu_\rho\; \delta^\nu_\sigma - \delta_\sigma ^\mu\; \delta _\rho ^\nu\Big]\bigg\}\;\omega(\gamma)^2\;.\label{40b}
	\end{align}
\end{subequations}
Here $R^{\mu \nu}{}_{\rho \sigma}$ is the Riemann tensor of the $d$-dimensional metric $g_{\mu \nu}^\text{R}(x^\rho)$. While we assume the latter to be Einstein, all other properties of $g_{\mu \nu}^\text{R}(x^\rho)$, in particular its Riemann tensor, are still completely unconstrained. In particular no assumptions about possible symmetries of $\Big(\mathscr{M}_d, \,g^\text{R}_{\mu \nu}\Big)$ have been made.

\bigskip
\noindent\textbf{(4) Maximum symmetry.} If $\Big(\mathscr{M}_d, \,g^\text{R}_{\mu \nu}\Big)$ happens to be maximally symmetric, its curvature tensor satisfies
\begin{equation}
	R^{\mu \nu}{}_{\rho \sigma} = H_\text{R}^2\; \Big[\delta^\mu_\rho\; \delta^\nu_\sigma - \delta_\sigma ^\mu\; \delta _\rho ^\nu\Big] \;,\label{41}
\end{equation}
and as a result, the components \eqref{40b} simplify correspondingly. In this case it is not difficult to see  that, in higher-dimensional language, the equations \eqref{40a} and \eqref{40b} with \eqref{41} read:
\begin{equation}
	{}^{(d+1)}R^{IJ}{}_{KL} = C\; d^{-1}\; \Big[\delta^I_K\; \delta^J_L - \delta_L ^I\; \delta _K ^J\Big] \;.
\end{equation}
This leads us to the following conclusion:

For every choice of $\big\{\varepsilon, \alpha_1, \alpha_2\}$ and of the $d$-dimensional Einstein metric $g^\text{R}_{\mu \nu}$, the $(d+1)$-dimensional metric  ${}^{(d+1)}g_{IJ}(\gamma, x^\mu)$ defined by eq.\eqref{d+1} is maximally symmetric if, and only if, $g^\text{R}_{\mu \nu}(x^\rho)$ is maximally symmetric. In this case, $g_{\mu \nu}^\text{R}$ has $\frac{1}{2}d(d+1)$ Killing vectors, while ${}^{(d+1)}g_{IJ}$ has $\frac{1}{2}(d+1)(d+2)$. 

Recall also that in the present paper we have fixed the sign of the cosmological constant from the outset: $\sigma \equiv \Lambda(k)/|\Lambda(k)| = +1$. Hence, in the maximally symmetric case, we are bound to consider families of spheres or de Sitter spacetimes,  $\mathscr{M}_d = \text{S}^d$ or \linebreak $\mathscr{M}_d = \text{dS}_d$, respectively, depending on whether the to-be-embedded manifolds are Euclidean or Lorentzian.

Furthermore, let us emphasize that this alternative, $\mathscr{M}_d$ being Euclidean or Lorentzian, did not get linked to the sign $\varepsilon$ in the course of the above calculations, neither by requiring ${}^{(d+1)}g_{IJ}$ to be Einstein, nor by demanding maximum symmetry.
\vspace{-0.cm}

\section{The candidates: AdS${}_{d+1}$ and dS${}_{d+1}$}\label{sec:3}
Let us return to the question raised in the Introduction: Which principles and criteria can constrain or, in the ideal case, determine uniquely a manifold $\mathscr{M}_{d+1}$ that geometrizes a given trajectory of Lorentzian spacetimes $\Big(\mathscr{M}_d, \,g^k_{\mu \nu}\Big)$?

\bigskip
\noindent\textbf{(1) Symmetry.} A natural principle of this kind, which we shall adopt here, is the following: The higher-dimensional $\mathscr{M}_{d+1}$ should display the maximum amount of symmetry that is consistent with the symmetry properties of the lower-dimensional metrics $g_{\mu \nu}^k$.

This principle unfolds its power in full if $\mathscr{M}_{d+1}$ can be required to be maximally symmetric. But, as we know, this will be possible only for generalized RG trajectories where already the original $\mathscr{M}_{d}$'s possess a corresponding symmetry.

Since it is our goal to find examples in which $\Big(\mathscr{M}_{d+1}, \,{}^{(d+1)}g_{IJ}\Big)$ is constrained as strongly as possible, we henceforth insist on maximally symmetric embedding manifolds $\mathscr{M}_{d+1}$.

To make sure that the latter can arise actually, we assume that, on all scales, $\Big(\mathscr{M}_d, \,g^k_{\mu \nu}\Big)$ is a maximally symmetric, and Lorentzian Einstein space with a positive cosmological constant. Hence from now on $\Big(\mathscr{M}_d, \,g^k_{\mu \nu}\Big)$ amounts to  de Sitter spacetimes dS${}_d$ with a running Hubble parameter $H = H(k)$.

The $k$-dependent effective field equation tells us that $g_{\mu \nu}^k \propto 1/\Lambda(k)$, and so we may write the running de Sitter metrics as follows [I]:
\begin{equation}
	g_{\mu \nu}^k \;=\; Y(k)^{-1}\;g^\text{R}_{\mu \nu} \qquad \text{where} \qquad Y(k) \;\equiv\; \frac{\Lambda(k)}{\Lambda_\text{R}}\;.
\end{equation}
This identifies the reference metric $g^\text{R}_{\mu \nu}$ used above  with the running metric $g_{\mu \nu}^k$ evaluated at some arbitrary $k \equiv k_\text{R}>0$.
\bigskip

\noindent\textbf{(2) AdS$_{d+1}$ and  dS$_{d+1}$ arise.} At this point, the basic problem has boiled down   to a question that we addressed already in the previous section, namely: Given a stack of de Sitter spaces $\Big(\mathscr{M}_d = \text{dS}_d, \,Y(k)^{-1}g_{\mu \nu}^\text{R}\Big)$, i.e., a family of spacetimes whose members are all \textit{Lorentzian} and \textit{maximally symmetric}, in which manifolds $\Big(\mathscr{M}_{d+1}, \,{}^{(d+1)}g_{IJ}\Big)$ can they  be embedded if we demand that the higher-dimensional scale-space-time, too, is Lorentzian and maximally symmetric?

The demand of being Lorentzian fixes the signature of $\mathscr{M}_{d+1}$ in the form \mbox{$(- +++\cdots)$}, thus avoiding the (exotic and potentially problematic) situation with two times, \linebreak\mbox{$(- -++\cdots)$}. Hence the only time direction of $\mathscr{M}_{d+1}$ is the one inherited from $\mathscr{M}_{d}$, and so \textit{the scale coordinate is determined to be a spatial one}.

As a consequence, we must set $\varepsilon=+1$ in our above catalog of possible local embeddings. 

The remaining freedom lies in the choice of the integration constants  $(\alpha_1, \alpha_2)$ then. It leaves us with only two principally different cases, namely $(\alpha_1, \alpha_2) = (1,0)$ and 	\linebreak $(\alpha_1, \alpha_2) = (0,1)$, respectively. According to \eqref{33}, the resulting cosmological constant of $\mathscr{M}_{d+1}$ is negative in the first, and positive in the second case:
\begin{empheq}[box=\fbox]{align}
	\mathbf{	{}^{(d+1)}\Lambda}{}_{(1,0)} &= -\frac{1}{2}\;d\;(d-1)\; H_\text{R}^2 = - \left(\frac{d}{d-2}\right) \;\Lambda_\text{R}\;,\\
	\mathbf{	{}^{(d+1)}\Lambda}{}_{(0,1)} &= +\frac{1}{2}\;d\;(d-1)\; H_\text{R}^2 = + \left(\frac{d}{d-2}\right)\; \Lambda_\text{R}\;.
\end{empheq}
Thus, insisting that $\mathscr{M}_{d+1}$ is maximally symmetric narrows down the possibilities to just two cases, namely \textit{$\mathscr{M}_{d+1}$ is either the anti-de Sitter spacetime} AdS${}_{d+1}$\textit{, or the de Sitter spacetime} dS${}_{d+1}$.
\bigskip

\noindent \textbf{(3) Scale coordinates $\gamma$ vs. $\xi$.} Our conclusions are easily checked explicitly on the basis of the conformal factors given by \eqref{24}:
\begin{equation}
	\Omega_{(1,0)}(\gamma) = 1/ \sinh(H_\text{R}\,\gamma)\;,
\end{equation}
\begin{equation}
	\Omega_{(0,1)}(\gamma) = 1/ \cosh(H_\text{R}\,\gamma)\;.
\end{equation}
\bigskip

\noindent\textbf{(3a)} In the $\mathbf{(1,0)}$ \textbf{case} the line element \eqref{d+1} assumes the following form:
\begin{equation}
	{}^{(d+1)}g_{IJ}^\text{AdS}(x^K) \di x^I \di x^J = \frac{1}{\sinh^2 (H_\text{R}\,\gamma)}\; \bigg[\,(\di \gamma)^2\; + \;\di \Sigma_d^2\;\bigg],\qquad \gamma \in (-\infty,\, 0)\;.\label{57}
\end{equation}
Here we introduced the special notation 
\begin{equation}
	\di \Sigma_d^2\; \equiv \;g_{\mu \nu}^\text{R} (x^\rho)\;\di x^\mu \di x^\nu
\end{equation}
for the metric of the $d$-dimensional de Sitter space dS$_{d}$ with the Hubble parameter $H_\text{R}$. By the coordinate transformation $\gamma \leftrightarrow\xi$ with
\begin{equation}
	\xi(\cdot):\; (-\infty,\;0)\;\to\; (0,\,\infty),\qquad \gamma\mapsto\xi(\gamma) = -H_\text{R}^{-1} \; \ln \; \tanh \left(-\frac{1}{2}\; H_\text{R}\, \gamma\right)
\end{equation}
the metric \eqref{57} can be brought to the alternative  form
\begin{equation}
	\boxed{	{}^{(d+1)}g_{IJ}^{\text{AdS}}(x^K) \di x^I \di x^J = (\di \xi)^2\; + \sinh^2 (H_\text{R}\,\xi)\;\di \Sigma_d^2\,,\qquad \xi \in (0,\,\infty)}\label{60}
\end{equation}
The line element \eqref{60} is known to describe a patch of the AdS${}_{d+1}$ manifold by slicing it with $d$-dimensional de Sitter spaces.\footnote{This AdS metric is used comparatively rarely; exceptions include \cite{Kumar, Karch}.}
\bigskip

\noindent\textbf{(3b)} 
In the $\mathbf{(0,1)}$ \textbf{case} we are similarly led to
\begin{equation}
	{}^{(d+1)}g_{IJ}^{\text{dS}}(x^K) \di x^I \di x^J = \frac{1}{\cosh^2 (H_\text{R}\,\gamma)}\; \bigg[\,(\di \gamma)^2\; + \;\di \Sigma_d^2\,\bigg]\,, \qquad \gamma \in (-\infty,\, \infty)
\end{equation}
which, by means of a different coordinate transformation,
\begin{equation}
	\xi(\cdot):\; (-\infty,\;\infty)\;\to\; (0,\,\pi \,H_{\text{R}}^{-1}),\qquad \gamma\mapsto	\xi(\gamma) = 2H_\text{R}^{-1} \;\arctan\left(e^{H_\text{R}\gamma}\right)\;,
\end{equation}
can be brought to the form
\begin{equation}
	\boxed{	{}^{(d+1)}g_{IJ}^{\text{dS}}(x^K) \di x^I \di x^J = (\di \xi)^2\; + \sin^2 (H_\text{R}\,\xi)\;\di \Sigma_d^2\,, \qquad H_\text{R} \,\xi \in (0,\,\pi)}\label{72}
\end{equation}
Eq.\eqref{72} is nothing but the well known metric of dS-sliced de Sitter space \cite{Griffiths}.

Note that in the above both $\gamma$ and $\xi$ have the dimension of a length, $[\gamma]=[\xi] = -1$, being the only exceptions to our convention that coordinates should be dimensionless.

\bigskip
\noindent \textbf{(4) (A)dS$_{d+1}$ cases combined.} It is convenient to combine the two relevant metrics, \eqref{60} and \eqref{72}, respectively, in the following fashion:
\begin{equation}
	{}^{(d+1)}g^\text{AdS/dS}_{IJ}(x^K) \di x^I \di x^J = (\di \xi)^2\; + F (H_\text{R}\,\xi)^2\;\di \Sigma_d^2\;.\label{75}
\end{equation}
Herein, the function $F$ is given by
\begin{equation}
	F(x) =\left\{ \begin{array}{ll}
		\sinh(x) \qquad \text{  for AdS}{}_{d+1} 
		\\	\sin(x) \qquad\;\,\, \text{  for dS}{}_{d+1} \qquad,
	\end{array}  
	\right.\ \label{3.13}
\end{equation}
and, as before, $\di\Sigma_d^2$ stands for a dS${}_d$ metric with the Hubble parameter $H_\text{R}$. 

\bigskip
\noindent \textbf{(5) Global coordinates on  dS$_{d}$.} So far the system of coordinates within the $\xi = const$ surfaces has been left unspecified. When explicit coordinates are needed, we shall choose  them in a way such that $x^\mu \equiv (t, \sigma^i)$, with spatial coordinates $\sigma^i$ and the time coordinate $t$, covers the maximal extension of  de Sitter space. In such global coordinates the dS${}_d$ metric reads \cite{Griffiths}:
\begin{equation}
	\di \Sigma_d^2 = \frac{1}{H_\text{R}^2} \; \bigg[-\di t^2 + \cosh^2 (t)\; \di \Omega^2_{d-1}\bigg]\;.\label{80}
\end{equation}
Here $\di \Omega^2_{d-1}$ denotes the line element of a unit $(d-1)$-sphere coordinatized by the $\sigma^i$'s.

\bigskip
\noindent \textbf{(6) Dimensionless scale coordinate \bm{$\bar \xi = H_\text{R} \xi$}.} Adopting the line element $\di \Sigma_d^2$ from eq.\eqref{80}, the combined metrics \eqref{75} assume the form
\begin{equation}
	{}^{(d+1)}g_{IJ}^\text{AdS/dS}(x^K)\di x^I \di x^J \;= \;H_\text{R}^{-2}\; \di s_{d+1}^2\Big|_{\bar \xi = H_R\, \xi}
\end{equation}
where $\di s_{d+1}^2$ denotes the dimensionless line element
\begin{equation}
	\boxed{\di s_{d+1}^2 = (\di \bar \xi)^2 + F(\bar \xi)^2\; \bigg[-\di t^2 + \cosh^2 (t)\; \di \Omega^2_{d-1}\,\bigg]\;.}\label{86}
\end{equation}
It depends on dimensionless coordinates only. They  include
\begin{equation}
	\bar \xi \equiv H_\text{R}\; \xi
\end{equation}
which labels the leaves of the foliation. This dimensionless scale coordinate amounts to the original one,  $\xi$, when expressed in units of the Hubble length \mbox{$L_H^\text{R} \equiv H_\text{R}^{-1}$} in the reference spacetime.

\section{Relating foliation and RG scale}\label{sec:4}
The spacetime $\Big(\mathscr{M}_{d+1}, \,{}^{(d+1)}g_{IJ}\Big)$  is foliated by leaves with $\xi =  const$ which we  would like to interpret as surfaces of equal RG scale $k$. By eq.\eqref{75}, our two candidates  $\mathscr{M}_{d+1} = \text{AdS}_{d+1}$  and $\mathscr{M}_{d+1} =$ dS${}_{d+1}$ induce certain metrics on those $d$-dimensional leaves.  We insist that these metrics  coincide exactly with those delivered by the renormalization group:
\begin{equation}
	{}^{(d+1)}g_{IJ}^{\text{AdS/dS}}\di x^I \di x^J \Big|_{\di \xi = 0} \;\;=\; \;F(H_\text{R}\,\xi)^2 \;\di \Sigma_d^2 \;\;\stackrel{!}{=}\; \;Y(k)^{-1} \; \di \Sigma _d^2
\end{equation}
This condition leads us to the following fundamental requirement for the viability of the suggested embeddings:
\begin{equation}
	\boxed{	F\left(H_\text{R}\, \xi\right) \;=\; Y(k)^{-1/2}\;, \qquad k \; \in \; \mathbb{R}^+}\label{101}
\end{equation}
The all-decisive question is whether eq.\eqref{101} gives rise to an acceptable relationship between $\xi$ and $k$, namely an admissible coordinate transformation $\xi = \xi(k)\leftrightarrow k = k(\xi)$.

Note that by virtue of
\begin{equation}
	Y(k) = \frac{\Lambda(k)}{\Lambda_\text{R}} =  \frac{H(k)^2}{H_\text{R}^2} \qquad \text{and} \qquad Y(k) ^{-1/2} =  \frac{H_\text{R}}{H(k)} = \frac{L_H(k)}{L_H^\text{R} }
\end{equation}
the requirement \eqref{101}, when expressed in terms of the respective Hubble lengths \linebreak$L_H(k) = 1/H(k)$ and $L_H^\text{R}=1/H_\text{R}$, writes
\begin{equation}
	\boxed{\hat F (\xi) = L_H(k) }\qquad \text{where} \qquad \hat F (\xi) \equiv L_\text{R} \;F\left(\frac{\xi}{L_H^\text{R}}\right)\;.\label{4.2/2}
\end{equation}
It can be observed that this condition is a 	``deformation'' of the one occurring in the Ricci flat case studied in [I]. There, the simpler condition $\xi = L_H(k)$  had appeared instead of \eqref{4.2/2}. However, as one might expect, at small arguments $\hat F$ approaches $\hat F(\xi) = \xi + O\Big((\xi/L_H^\text{R})^3\Big)$ for the 	``deformed'' functions of \eqref{3.13}, $\hat F (\xi) = L_\text{R}\sin(\text{h})\Big(\xi/L_\text{R}\Big)$.

In the sequel we investigate the properties of the  relationship \eqref{101} for the most interesting case of $d = 4$, i.e., the embedding of 4-dimensional scale dependent spacetimes into a 5-dimensional manifold $\mathscr{M}_5$.

In the rest of this paper, in discussions of a general nature that do not depend on the input from the RG flow, we shall continue to leave the dimensionality $d$ arbitrary, however.

\subsection{Information  from the RG}
To decide about the viability of an embedding, essential use must be made of the properties of the function $Y(k)^{-1/2} \equiv \sqrt{\Lambda_\text{R}/\Lambda(k)}$. The latter is determined by the RG trajectories of the Einstein-Hilbert truncation, concretely those of Type IIIa if $\Lambda_0 > 0$, and of Type IIa for $\Lambda_0 = 0$, respectively. Their investigation by both analytical and numerical means has revealed the following properties that are going to be relevant \cite{Martin, Frank1, Renata1}:
\bigskip

\noindent\textbf{Case \bm{$\Lambda_0 >0$}:} The function $Y(\cdot)^{-1/2}: \mathbb{R}^+\to \mathbb{R}^+, \; k \mapsto Y(k)^{-1/2}$ is a smooth, strictly decreasing function which maps the $k$-interval $(0, \infty)$ invertibly on the interval $(0, y^{-1})$.

Here we introduced the parameter $y^2\equiv Y(0)=\Lambda_0/\Lambda_\text{R}>0$, or equivalently, \linebreak$y = H(0)/H_\text{R} = L^\text{R}_H/L_H(0)$, employing the standard definitions $L_H (k) \equiv 1/H(k)$ and $L^\text{R}_H \equiv 1/H_\text{R}$.
\bigskip

\noindent\textbf{Case \bm{$\Lambda_0 =0$}:} The function $Y(\cdot)^{-1/2}: \mathbb{R}^+\to \mathbb{R}^+, \; k \mapsto Y(k)^{-1/2}$ is a smooth, strictly decreasing function  which maps the $k$-interval $(0, \infty)$ invertibly on the interval $(0, \infty)$. 

\bigskip
In either case, the domain and the codomain of $Y(\cdot)^{-1/2}$ are always $\mathbb{R}^+$. The two cases differ however with respect to the actual image of $\mathbb{R}^+$ under $Y(\cdot)^{-1/2}$: If $\Lambda_0 > 0$, the image consists of the interval $(0, y^{-1})$ only, while it comprises all of $\mathbb{R}^+$ for $\Lambda_0=0$. This is related to the fact that $Y(k)^{-1/2}$ approaches a finite limit $\lim_{k\to 0}Y(k)^{-1/2} =y^{-1}$ when $\Lambda_0 >0$, but it diverges to $+\infty$ if we let $k \to 0$ for $\Lambda_0 = 0$.

\begin{figure}[t]
	\centering
	\includegraphics[scale=0.4]{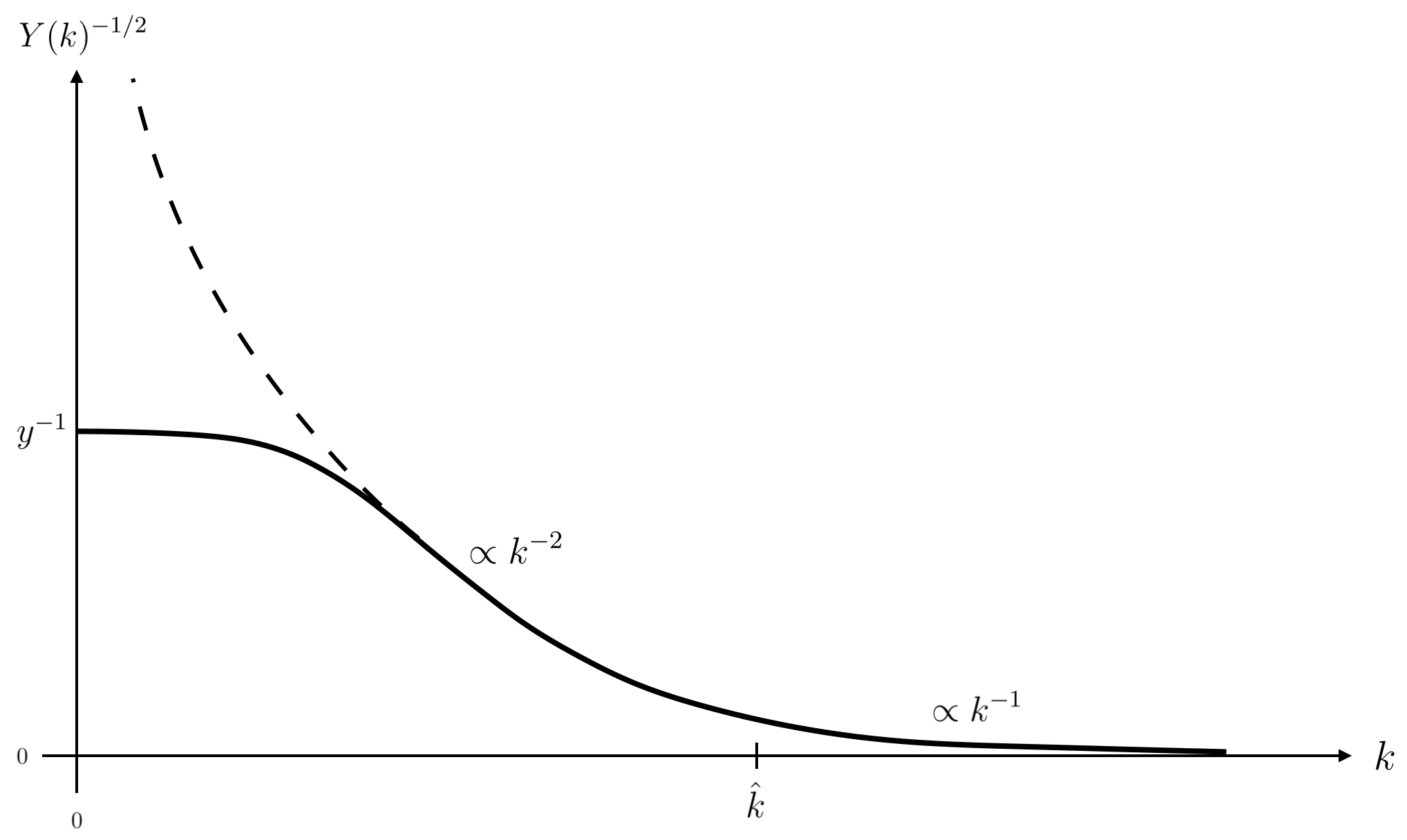}
	\caption{The schematic behavior of the function $Y(k)^{-1/2}$ for $\Lambda_0>0$ (solid line) and $\Lambda_0=0$ (dashed line).}\label{Y-schematic}
\end{figure}

In Figure \ref{Y-schematic} we schematically depict the behavior of $Y(k)^{-1/2} \equiv Z(k)$ in the two cases.
The relevant characteristics of this function are also well described by the following analytic approximation [I]:
\begin{equation}
	Y(k)^{-1/2}\;\approx\;\;\left\{ \begin{array}{ll}
		\Big[y^2 +(\ell\;k)^4\Big]^{-1/2}\qquad \text{  for } \quad 0\leq k \lesssim\hat k\\
		(L\;k)^{-1} \qquad \qquad \qquad\text{ for }\quad  \hat k \lesssim k <\infty\;.
	\end{array}  
	\right.\label{4.3}
\end{equation}
Details concerning the approximate formula for $\Lambda(k)$ that underlies eq.\eqref{4.3} can be found in the Appendix.

It is obvious from  Figure \ref{Y-schematic} that, for $\Lambda_0>0$, $Z(k) = Y(k)^{-1/2}$ possesses a monotone inverse function $Z^{-1}: \left(0,y^{-1}\right) \to \mathbb{R}^+$, and that the image of $\left(0,y^{-1}\right)$ under $Z^{-1}$ is the full $\mathbb{R}^+$, i.e., the complete half-line of scales $k$. If instead $\Lambda_0 = 0$, the inverse is a certain function $Z^{-1}: \mathbb{R}^+ \to \mathbb{R}^+$, but again, $Z^{-1}$ is monotone and has an image that covers the entire $\mathbb{R}^+$ of $k$-values.

\subsection{The $k$-$\xi$ transformation: the AdS${}_{5}$ candidate}
As a step towards establishing the viability of the AdS${}_{d+1}\equiv\text{AdS}_5$ embedding, let us now try to satisfy the requirement
\begin{equation}
	\boxed{\sinh \left(H_\text{R}\, \xi\right) \;=\; Y(k)^{-1/2}\; \equiv \;Z(k)}\label{120}
\end{equation}
by some monotone function $\xi = \xi(k)$, or $k = k(\xi)$, respectively.

\bigskip
\noindent \textbf{(1) Existence.} Solving the relation \eqref{120} for $\xi$ yields a function  which is well defined for all $k \in \mathbb{R}^+$, and assumes values in $\mathbb{R}^+$:
\begin{equation}
	\xi(\cdot):\, \mathbb{R}^+ \to \mathbb{R}^+,\qquad k \to\xi(k) \;= \;H_\text{R}^{-1}\; \text{arsinh}\bigg(Y(k)^{-1/2}\bigg)\;.\label{121}
\end{equation}
The key observation in the previous subsection, namely that $Y(\cdot)^{-1/2}$ has an everywhere negative derivative, for both $\Lambda_0 =0$ and $\Lambda_0 >0$, implies that the function \eqref{121} is strictly decreasing, too:  $\frac{d}{dk} \xi(k) < 0$ for all $k > 0$. This is precisely as it must be if the $k$-$\xi$ relationship is to qualify as an orientation reversing\footnote{The orientation reversing character of the coordinate transformation $k \to \xi$ is a trivial consequence of $\xi$ being a \textit{length}, while $k$ is a \textit{momentum}.} diffeomorphism.

As for the inverse map $\xi \mapsto k(\xi)$, we solve \eqref{120} for $k$ this time:
\begin{equation}
	k (\xi) \;=\; Z^{-1}\bigg(\sinh \big(H_\text{R}\,\xi\big)\bigg)\;.\label{125}
\end{equation}
The monotonicity of $Z^{-1}(\cdot)$ and $\sinh (\cdot)$ implies that, as expected, $\frac{d}{d\xi}k(\xi) < 0$ everywhere.  From the properties of $Z^{-1}$ we furthermore infer the domain of the function defined by the expression \eqref{125}:
\begin{eqnarray}
	&k(\cdot):\;  \Big(0,\; \xi_\text{max}(y)\Big)\; \to \; \mathbb{R}^+, \quad &\xi\; \mapsto \;k(\xi)\qquad \text{for } \;\;\Lambda_0 > 0\\
	&k(\cdot): \; \mathbb{R}^+\; \to \; \mathbb{R}^+, \qquad\qquad\quad \;&\xi\; \mapsto \;k(\xi)\qquad \text{for } \;\,\Lambda_0 = 0
\end{eqnarray}
Here we introduced the $y$-dependent interval boundary
\begin{equation}
	\xi_\text{max} (y) \; \equiv\; H_\text{R}^{-1}\;\; \text{arsinh} \big(y^{-1}\big)\;.\label{128}
\end{equation}

Since the image of $k(\cdot)$ equals $\mathbb{R}^+$ in both cases, we can conclude that it is sufficient to draw $\xi$-values from the interval $(0, \,\xi_\text{max})$ in the first, and from $\mathbb{R}^+$ in the second case, in order to parameterize all scales $k \in \mathbb{R}^+ \equiv (0, \,\infty)$ in a smooth and invertible manner.

This proves the existence of an admissible transformation $\xi \leftrightarrow k$  which possesses the desired properties of a one-dimensional diffeomorphism. It entails the viability of an embedding into the 5-dimensional anti-de Sitter space AdS$_5$, and this is what we wanted to establish.
\bigskip

\noindent \textbf{(2) Asymptotic form.} Within the approximation of eq.\eqref{4.3}, we can write down the coordinate transformation relating $\xi$ and $k$ in closed form. In the asymptotic scaling regime near the UV fixed point, $k \gtrsim\hat k$, we obtain for the Type IIIa and IIa trajectories alike:
\begin{equation}
	\xi (k) \;\approx\; H_\text{R}^{-1} \; \text{arsinh}\left(\frac{1}{L\,k}\right) \; \approx\; \frac{1}{H_\text{R}\,L\,k}
\end{equation}
Using the definitions of $H_\text{R}$ and $L$ in $d = 4$, this relation becomes
\begin{equation}
	\boxed{	\xi (k) \;\approx\; \left(\frac{\lambda_\ast}{3}\right)^{1/2}\; \frac{1}{k}\;.}
\end{equation}
It expresses a perfect inverse proportionality between the RG parameter $k$ and the value of the (dimensionful) coordinate $\xi$.

In the semiclassical regime  $0 <k\lesssim\hat k$, the properties of the corresponding transformation
\begin{equation}
	\boxed{	\xi (k) \;\approx\; H_\text{R}^{-1} \; \text{arsinh}\Big(\big[y^2+(\ell\,k)^4\big]^{-1/2}\Big) }
\end{equation}
depend on the type of the RG trajectory in a significant way.

For  Type IIIa trajectories, $y^2 \equiv \Lambda_0/\Lambda_\text{R}$ is non-zero, and  the $\xi$ coordinate corresponding to the limit $k\searrow  0$ is \textit{finite}:
\begin{equation}
	\xi (k = 0) \;\approx\; H_\text{R}^{-1} \; \text{arsinh}\big(y^{-1}\big) \;.
\end{equation}
For the Type IIa trajectory, on the other hand, $y = 0$ implies the following behavior of $\xi$ at low RG scales $k \ll \ell^{-1}$:
\begin{equation}
	\xi (k) \;\approx\; H_\text{R}^{-1} \; \text{arsinh}\left(\frac{1}{\ell^2\,k^2}\right) \; \approx\;  H_\text{R}^{-1}\;\ln \left(\frac{2}{\ell^2\,k^2}\right)\;.
\end{equation}
Hence $\xi(k)$ approaches $+\infty$ when $k$ decreases towards zero.

\subsection{The $k$-$\xi$ transformation: the dS${}_{5}$ candidate} \label{sec:4.3}
Let us return to the fundamental requirement \eqref{101} with a generic function $F(\cdot)$ appearing on its LHS. Since, first, $Y(k)^{-1/2}$ was found to be monotone, and second, $k = k(\xi)$ was demanded to be monotone, it follows that the requirement \eqref{101} can be satisfied only if  $F(H_\text{R}\xi)$ on its LHS has a monotone dependence on $\xi$.

Above, in the AdS case, this has indeed been the case, thanks to the monotonicity of the $\sinh$-function in eq.\eqref{120}.

For the second candidate, the de Sitter space dS${}_{d+1}$, the  requirement reads
\begin{equation}
	\boxed{\sin(H_\text{R}\,\xi) \;=\; Y(k)^{-1/2}\;\equiv\; Z(k)\;,}\label{I2-6}
\end{equation}
and here the situation is different, for two reasons:

\bigskip
\noindent(a) The line element \eqref{72} is non-degenerate only for $H_\text{R}\xi \in (0,\pi)$. Therefore, right from the outset we must restrict the range of the $\xi$ coordinate to the interval $\xi \in (0, \pi)\,H_\text{R}^{-1}$. 

\bigskip
\noindent (b) On this latter interval, the LHS of \eqref{I2-6}, $\sin(H_R\xi)$, is \textit{not} a monotone function of $\xi$. As a way out, we restrict $\xi$ even further, namely to only half of the original range:
\begin{equation}
\boxed{	\xi \;\in\; \left(0, \;\frac{\pi}{2}\right)\;H_\text{R}^{-1}\;.}\label{I2-7}
\end{equation}

For coordinate values in the interval \eqref{I2-7}, the metric \eqref{72} is well defined, and at the same time the LHS of \eqref{I2-6} is monotone with respect to $\xi$. Hence we can hope to find a diffeomorphism relating $\xi $ to $k$.

A discussion analogous to the one above shows that there is indeed such a coordinate transformation, albeit only for certain values of $y$.

\bigskip
\noindent\textbf{(1) Existence for $\bm{y>1}$.} Clearly, when it exists, the transformation has the form 
\begin{equation}
	\xi(k) = H_\text{R}^{-1} \arcsin \Big(Z(k)\Big) \quad \Longleftrightarrow\quad k(\xi) = Z^{-1} \Big(\sin(H_\text{R}\xi)\Big)	\label{4.50}
\end{equation}
and every RG scale $k \in \mathbb{R}^+$ gets related in a 1-1 way to a unique $\xi \in \big(0, \,\xi_\text{max}(y)\big)$ whereby 
\begin{equation}
	\xi_\text{max} (y) \; =\; H_\text{R}^{-1}\;\; \text{arcsin} \big(y^{-1}\big)\;.\label{4.51}
\end{equation}

For the expressions in \eqref{4.50} to make sense, the argument of the arcsin-function must satisfy $Z(k) \equiv Y(k)^{-1/2} \in (0,1)$ for all $k \in \mathbb{R}^+$. A quick glance at Figure \ref{Y-schematic} reveals that this is the case if, and only if $y >1$. Recalling the definition of the parameter $y$,
\begin{equation}
	y \equiv Y(0)^{1/2}\; = \;\Big(\Lambda_0/\Lambda_\text{R}\Big)^{1/2}\; =\; H(0)/H_\text{R}\;=\; L_H^\text{R}/L_H(0)\;,\label{4.52}
\end{equation}
we see that $y >1$ imposes a constraint on the value of the Hubble length at the trajectory's endpoint, $L_H(0) \equiv \lim_{k \to 0} L_H(k)$, in relation to the Hubble radius of the reference metric:
\begin{equation}
\boxed{	y>1 \quad \Longleftrightarrow\quad L_H^\text{R} >L_H(0) \quad \Longleftrightarrow\quad \Lambda_0 >\Lambda_\text{R}\;.}\label{4.53}
\end{equation}

If the  constraint \eqref{4.53} is satisfied, \eqref{4.50} does indeed define a diffeomorphic map $\xi(\cdot):\, \mathbb{R}^+\to \big(0,\,\xi_\text{max}\big)$, $k \mapsto\xi(k)$, as it is necessary for the embedding in dS${}_5$ to exist. In the opposite case, $y < 1$, no such map exists.\footnote{This is obvious from the relation \eqref{I2-6} already: Its RHS, $Y(k)^{-1/2}$, assumes values between zero and $y^{-1}>1$, while the magnitude of the LHS, $\sin(H_R\xi)$, never exceeds unity.}
\bigskip

\noindent\textbf{(2) Asymptotic form.} For the $k$-$\xi$ relationship at asymptotically large and small RG parameters, the analytic approximation \eqref{4.3} yields, respectively,
\begin{eqnarray}
	\xi (k)\approx& \left(\ffrac{L_{H}^\text{R}}{L}\right)\ffrac{1}{k}\quad\qquad\qquad\qquad\qquad\qquad\qquad&(L\;k\to \infty)\;,\\
	\xi (k)\approx& \xi_\text{max} (y)-\ffrac{L_H^\text{R}\ell^4}{2y^2 \sqrt{y^2-1}}\;k^4\qquad\qquad&(\ell \;k \to 0)\;.\label{4.61}
\end{eqnarray}
The $y$-dependence in the  prefactor of the $k^4$ term in \eqref{4.61} makes it quite clear that $y = 1$ amounts to a threshold that cannot be crossed. We shall come back to it towards the end of the next section.

\section{Global structure and (A)dS connections}\label{sec:5}

In the previous section we saw that $\xi$, the coordinate that labels the leaves of the foliation, and $k$, the RG scale, are indeed related by an admissible coordinate transformation. The next question we must address is how much of the total (A)dS${}_{d+1}$ manifold is actually covered by the embedded spacetimes $\bigg\{\Big(\mathscr{M}_{d}, \,g_{\mu \nu}^k\Big),\; k \in \mathbb{R}^+\bigg\}$ from the renormalization group. We consider the cases $\mathscr{M}_{d+1}=\text{AdS}_{d+1}$ and $\mathscr{M}_{d+1}=\text{dS}_{d+1}$ in turn.

\subsection{The AdS embedding}
We start by discussing the geometrization of the RG flow by means of $\mathscr{M}_{d+1}=\text{AdS}_{d+1}$. According to eq.\eqref{86}, the dS-sliced AdS metric reads, in dimensionless form:\footnote{For simplicity, we omit the overbar from $\bar \xi$ in this subsection, i.e., now $\xi$ is understood to be measured in units of the reference Hubble length $H_\text{R}^{-1}$.}
\begin{equation}
	\di s_{d+1}^2 = (\di  \xi)^2 + \sinh^2( \xi)\; \bigg[-\di t^2 + \cosh^2 (t)\; \di \Omega^2_{d-1}\,\bigg]\label{120bis}
\end{equation}
If we leave the relation of  $\xi$ to $k$ aside for a moment, the maximal range of the coordinate values at which \eqref{120bis} can be applied is given by
\begin{equation}
	t\;\in\; \big(-\infty,\; +\infty\big) \qquad \text{ and }\qquad \xi \;\in\; \big(0,\; \infty\big)\;.\label{121bis}
\end{equation}

\bigskip
\noindent \textbf{(1) Relationship $\xi$-$k$ disregarded.} To find out how \eqref{120bis} is connected to the global AdS spacetime, and also in order to derive the corresponding Penrose diagram, let us perform the coordinate transformation $(\xi,\,t) \to (r,\,\tau)$ given by
\begin{eqnarray}
	r\;&=&\; \sinh(\xi) \;\;\cosh(t)\;,\nonumber\\
	\tan(\tau)\;&=&\; \tanh(\xi)\;\; \sinh(t)\;.
\end{eqnarray}
This transformation turns   the line element \eqref{120bis} into
\begin{equation}
	\di s_{d+1}^2 = -\Big(1+r^2\Big)\; \di \tau^2 \;+\; \frac{\di r^2}{1+r^2} \;+\; r^2\; \di \Omega_{d+1}^2\label{123}
\end{equation}
which is the well known AdS${}_{d+1}$ metric in \textit{global coordinates}, and for a unit Hubble parameter. The latter metric can be applied for the coordinate ranges
\begin{equation}
	\tau\;\in\; \big(-\infty,\; +\infty\big) \qquad \text{and}\qquad r \;\in\; \big(0,\; \infty\big)\;,
\end{equation}
\\

\bigskip
\begin{wrapfigure}{l}{0.48\textwidth}
	\centering
	\includegraphics[scale=0.61]{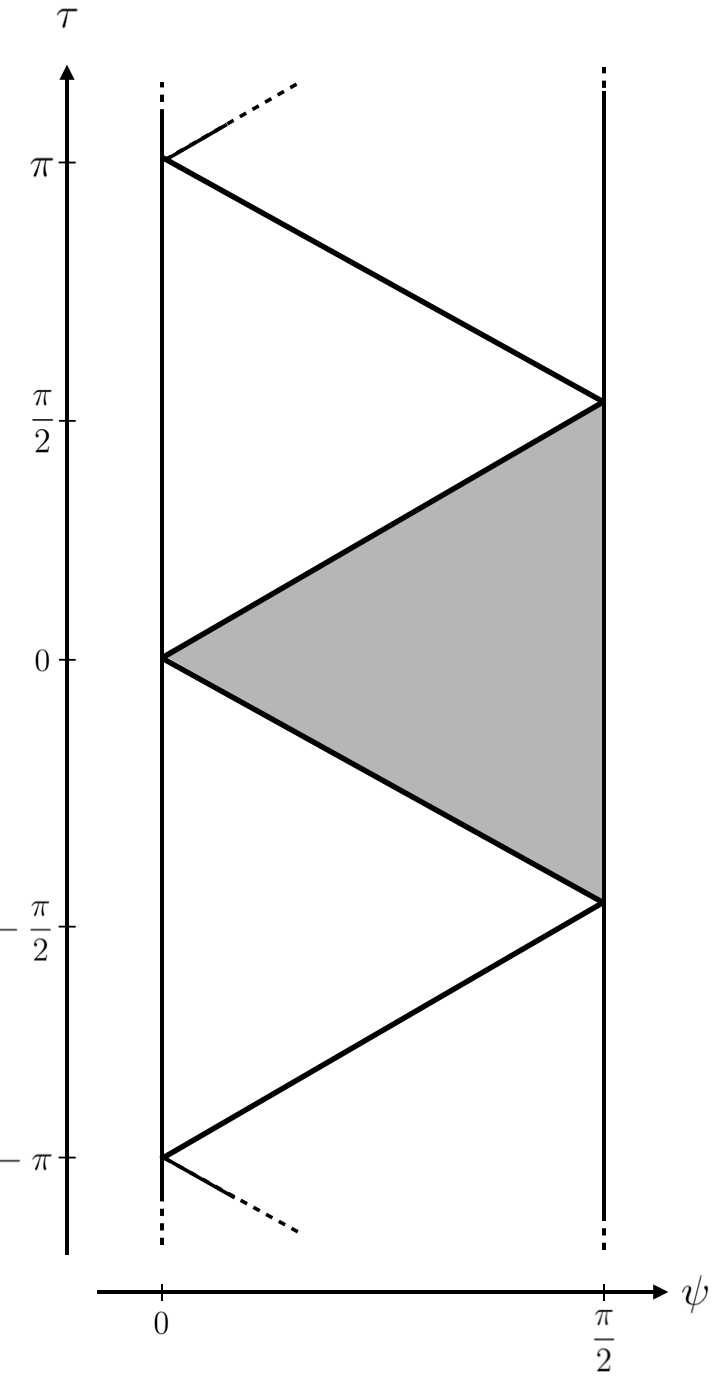}
	\caption{The Penrose diagram of (the universal cover of) the AdS${}_{d+1}$ spacetime. The shaded triangle corresponds to the part where the metric \eqref{120bis} applies when its maximum coordinate range \eqref{121bis} is exploited.\vspace{1cm}}\label{Penrose-cover}
\end{wrapfigure}
\noindent which actually correspond to the universal cover of the AdS${}_{d+1}$ spacetime.

The causal structure of the global manifold becomes manifest after an additional coordinate transformation $r \to \psi$:
\begin{equation}
	\tan (\psi)\;=\; r\;.
\end{equation}
It  converts \eqref{123} to the manifestly conformally flat metric
\begin{equation}
	\di s_{d+1}^2 = \frac{1}{\cos^2(\psi)}\bigg[-\di \tau^2+ \di \psi^2+ \sin^2(\psi)\; \di \Omega^2_{d+1}\bigg].\label{126}
\end{equation}
This metric is applicable for $\tau \in (-\infty, \infty)$, \mbox{$ \psi \in \left(0, \frac{\pi}{2}\right)$}.

\begin{figure}[t]
	\centering
	\includegraphics[scale=0.49]{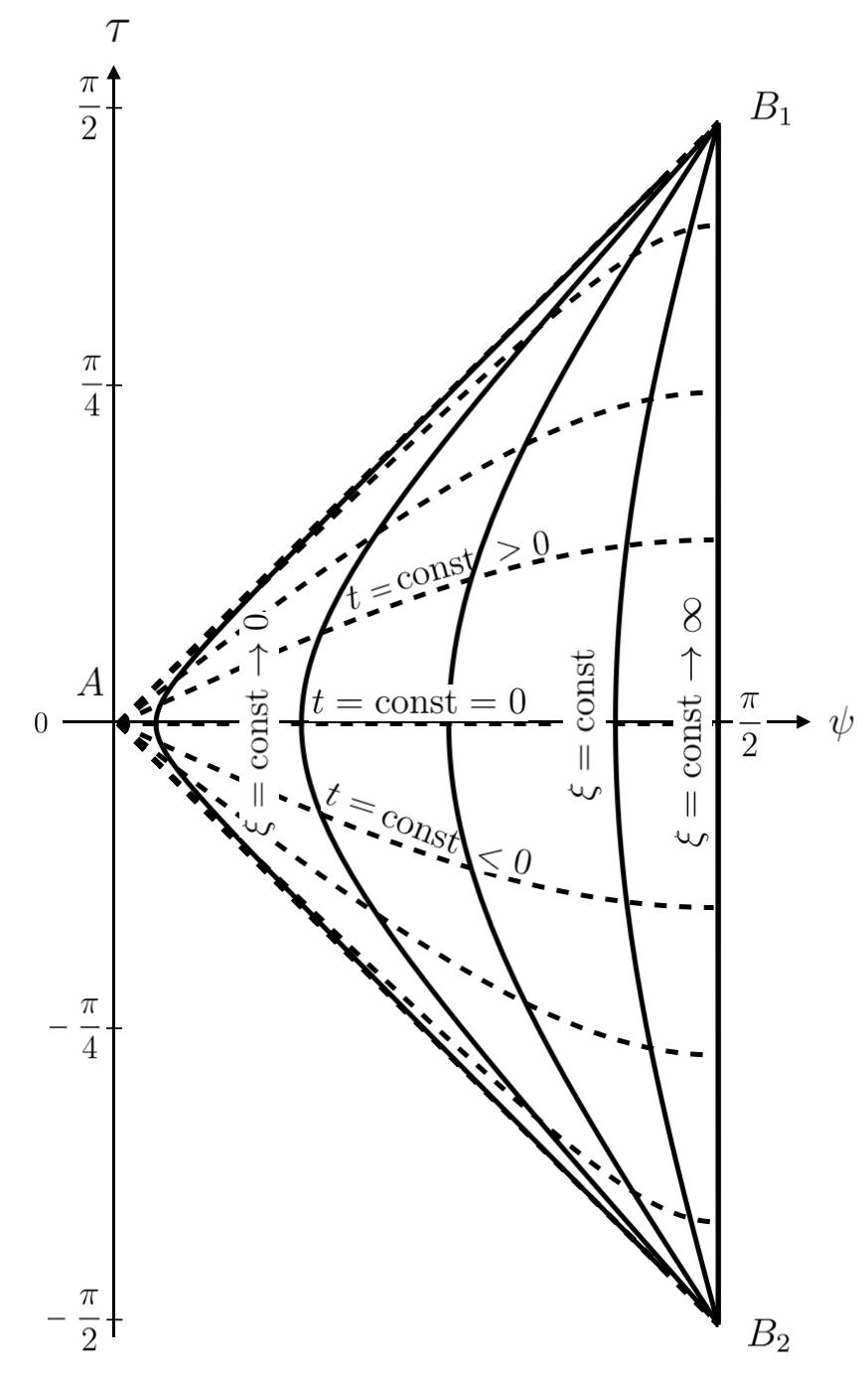}
	\caption{The shaded triangle of the AdS${}_{d+1}$ Penrose diagram in Fig.\ref{Penrose-cover} is redrawn. On the $\tau$-$\psi$ plane, various coordinate lines are shown on which $\xi = const$ (solid lines) or $t = const$ (dashed lines). In the limits $\xi = const \to \infty$, and $\xi =  const \to 0$, the surfaces of constant scale are seen to approach the line segment $B_1B_2$, and the null cone $B_1AB_2$, respectively.}\label{Penrose-coord}
\end{figure}

In Figure \ref{Penrose-cover} we sketch the Penrose diagram obtained from \eqref{126} after multiplication by $\cos^2(\psi)$. In this diagram, every point on the $\tau$-$\psi$ plane corresponds to a sphere S${}^{d-1}$ with radius $\sin^2(\psi)$.

It can be checked that the original metric \eqref{120bis} with coordinate ranges \eqref{121bis} covers only a part of the entire manifold. In Figure \ref{Penrose-cover} the corresponding patch is indicated by the shaded triangle.

Furthermore, in Figure \ref{Penrose-coord}, various coordinate lines having constant $t$- or $\xi$-values are shown within this triangular region. Every line with $\xi = const$ represents one of the $d$-dimensional spacetimes $\Big(\text{dS}_d, \,g^{k}_{\mu \nu}\Big)\Big|_{k = k(\xi)}$ which were supplied by the renormalization group.

\bigskip

\noindent \textbf{(2) Relationship $\xi$-$k$ imposed, Type IIIa.} In Section \ref{sec:4} we concluded that, if $\Lambda_0 \neq 0$, the scale coordinate $\xi = \xi(k) \in (0,\,\xi_\text{max})$ does not exhaust the full range of theoretically possible values, $\xi \in (0, \infty)$:
\begin{eqnarray}
&	\Bigg\{\big(\text{dS}_d, \;g_{\mu \nu}^k\big), \; k \in \mathbb{R}^+\Bigg\} &\;=\; 	\Bigg\{\big(\text{dS}_d, \;g_{\mu \nu}^{k(\xi)}\big), \;\xi \in \big(0,\; \xi_\text{max}(y)\big)\Bigg\}\;.\label{5.all}
\end{eqnarray}

As a consequence, the entirety of all spacetimes  that occur along a complete Type IIIa trajectory require for their embedding  only a part of the triangular region in the $\tau$-$\psi$ plane. It is given by the shaded area of the Penrose diagram in Figure \ref{Penrose-IIIa}. This area is coordinatized by $\xi \in \left(0,\,\xi_\text{max}(y)\right)$,  $t \in (-\infty, \infty)$.

\bigskip
\noindent\textbf{(3) Relationship $\xi$-$k$ imposed, Type IIa.} If $\Lambda_0=0$ on the other hand, i.e., for the Type IIa trajectory, the complete shaded triangle of Figure \ref{Penrose-cover}, but not more than that, is needed in order to embed the entire stack of spacetimes \mbox{$\bigg\{\Big(\text{dS}_d, \,g^{k}_{\mu \nu}\Big), \; k \in \mathbb{R}^+\bigg\}$}. See Figure \ref{Penrose-IIa} for an illustration of this case.

\subsection{Summary of the AdS case and AdS/CFT interpretation} \label{sec:5.2}

\begin{wrapfigure}{l}{0.4\textwidth}
	\centering
	\includegraphics[scale =0.5]{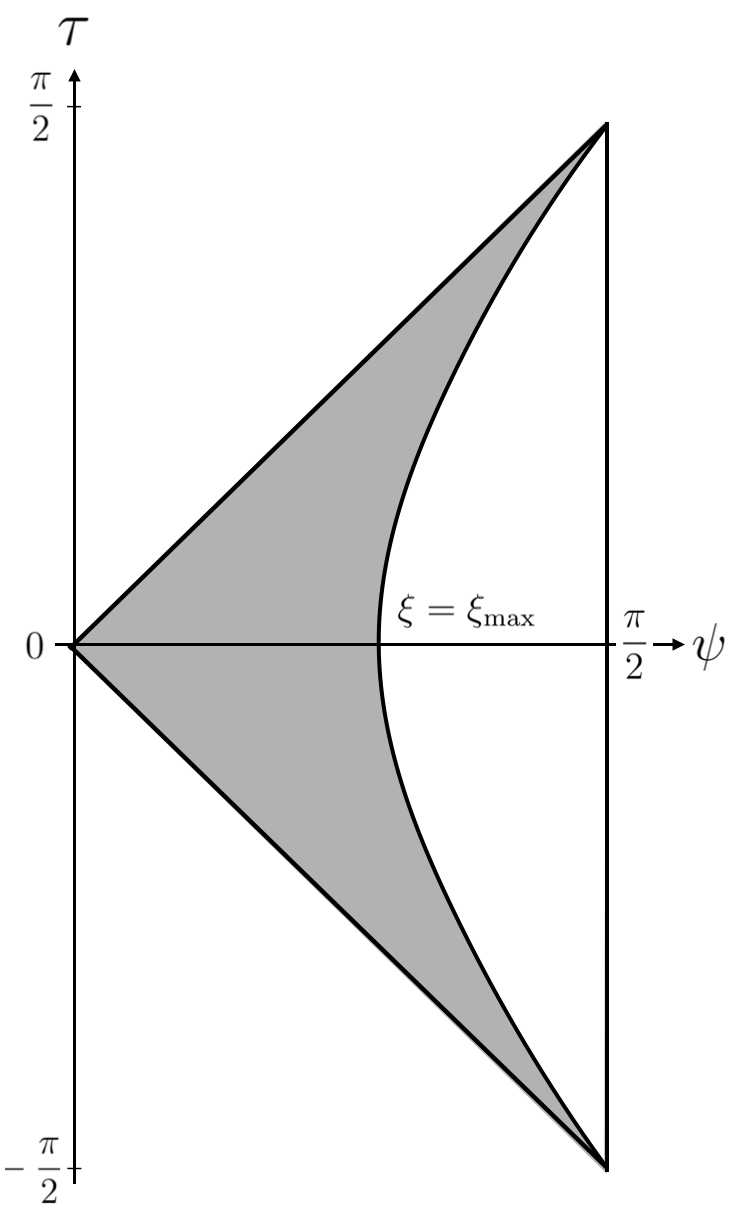}
	\captionof{figure}{The shaded area in this diagram indicates the part of the AdS$_5$ Penrose diagram in Figures \ref{Penrose-cover} and \ref{Penrose-coord} that is needed in order to embed all 4D spacetimes occurring along an RG trajectory of Type IIIa. The area is bounded by the $\xi=const$ line with $\xi =\xi_\text{max}(y)$ given in eq.\eqref{128}.\vspace{0.5 cm}}
	\label{Penrose-IIIa}
\end{wrapfigure}
We started out from a stack of de Sitter spacetimes which emerged as solutions to the scale dependent effective field equations derived from Type IIIa and IIa trajectories of running actions, $k \mapsto \Gamma_k$, $k \in \mathbb{R}^+$. We tried to interpret them as the leaves of a $(d+1)$-dimensional manifold that carries a natural foliation induced by the RG scale. We demanded that its metric ${}^{(d+1)}g_{IJ}$ should, (i), be Einstein and Lorentzian, (ii), have vanishing shift vector and $x^\mu$-independent lapse when presented in 	``scale-ADM'' form, and (iii), possesses as many Killing vectors as it is compatible with the other requirements and the symmetries of the input metrics $g_{\mu \nu}^k$.

We have then shown that, besides the dS candidate to be discussed below, an embedding into AdS$_d$ is the only option. Moreover, at least for $d = 4$, the properties of the RG flow are indeed such that the viability of the program can be demonstrated, i.e., there exists a scalar function $k = k(x^I)$ on the embedding manifold which describes the foliation by leaves of constant scale in a (in principle) coordinate independent manner.

\begin{figure}[t]
	\centering
	\includegraphics[scale =0.5]{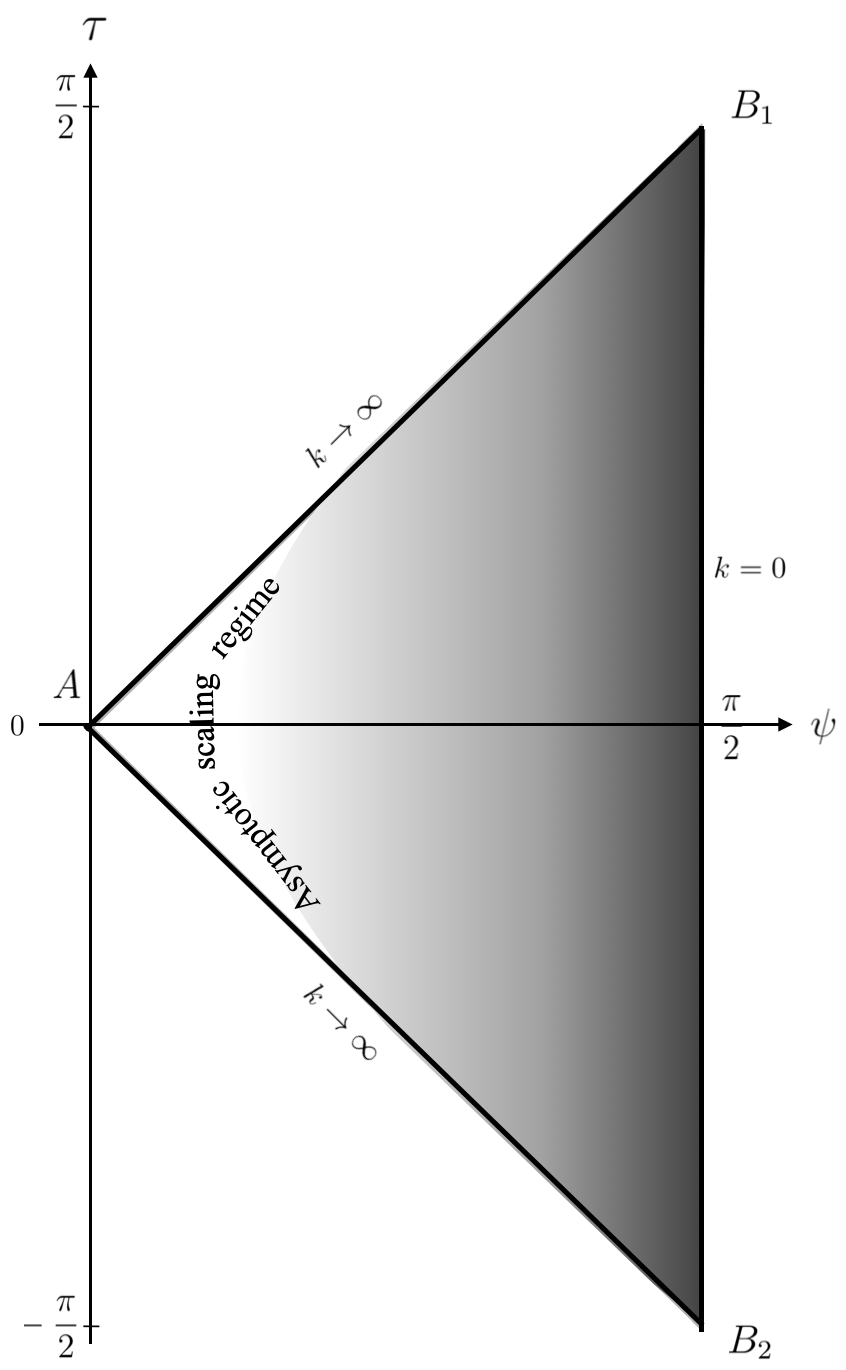}
	\captionof{figure}{Penrose diagram illustrating the portion of the AdS${}_5$ manifold that is necessary to embed all dS${}_4$ spacetimes along the Type IIa trajectory. The gray shading indicates the local value of $k$. Dark (light) regions correspond to low (high) values of the RG parameter $k$.}
	\label{Penrose-IIa}
\end{figure}

We demonstrated that  embedding the set  of all spacetimes \eqref{5.all} does not exhaust the entire AdS$_5$ manifold. The part of the latter which  actually comes into play is represented pictorially in the causal diagrams of Figures \ref{Penrose-IIIa} and \ref{Penrose-IIa}. They apply to 4D cosmological constants $\Lambda_0 > 0$ and $\Lambda_0 = 0$, respectively.

As for the interpretation, let us begin with the case $\Lambda_0 = 0$, i.e., the Type IIa trajectory. It is special in that it terminates in the Gaussian fixed point, and that $\lim_{k \to 0} \Lambda (k)=0$.\footnote{For recent work that independently hints at a special status of the Type IIa trajectory see \cite{Kevin, Max1, Max2}.} As a result, at the endpoint of the IIa trajectory the solution of the field equation changes from dS$_4$ to Minkowski space.

\bigskip
\clearpage
\noindent \textbf{(1) Geometry.} In Figure \ref{Penrose-IIa}, the triangle represents the part of AdS$_5$ covered by the embedding, and its causal properties in a global fashion. We denote this part as AdS$_5^\text{emb}$ in the following. We also recall that all points of the diagram are actually representatives of 3-dimensional spatial spheres.\\

In Figure \ref{Penrose-IIa}, the $\tau$-$\psi$ projection of the boundary of  AdS$_5^\text{emb}$ is seen to comprise 3 components: the lightlike ones $AB_1$ and $AB_2$, and a timelike one, $B_1B_2$, residing at spatial infinity.  While the line segments $AB_1$ and $AB_2$ consist of inner points of the full AdS$_5$ manifold\footnote{except for the points $A$, $B_1$, and $B_2$.}, the segment $B_1B_2$ is a boundary of both AdS$_5^\text{emb}$ and the total AdS$_5^\text{}$.\\

\bigskip
\clearpage

\noindent \textbf{(2) Quantum field theory à la GEAA.} Now let us switch from geometry to quantum field theory (QFT), formulated in GEAA language. For this purpose we decorated the  Penrose diagram in Figure \ref{Penrose-IIa} with information concerning the local value of $k(x^I)$ by means of a variable gray shading. It highlights the following properties of the foliation carried by  AdS$_5^\text{emb}$:
\bigskip

\noindent \textbf{(i)} In the UV limit $k \to \infty \Leftrightarrow \xi \to 0$, the leaves of constant scale  which describe 4D spacetime at a given resolution, get squeezed into the null cone $B_1AB_2$: the smaller is $\xi$, the more the lines with $\xi = const$ approach the diagonals $AB_{1,2}$ in the $\tau$-$\psi$ plane, see Figure \ref{Penrose-coord}. As a result, those  particular (spacelike!) leaves whose internal dynamics is ruled by the GEAA in the \textit{``bare'' limit}, $\Gamma_{k \to \infty} \sim S$, are situated infinitely close to the null cone $B_1AB_2$, the lightlike boundary of  AdS$_5^\text{emb}$.

\bigskip

\noindent \textbf{(ii)} Conversely,  in the \textit{physical limit} $k \to 0$, i.e., when the IR cutoff is removed, the 4D dynamics is governed by the ordinary effective action $\lim_{k \to 0}\Gamma_k = \Gamma$. This is the relevant action for the effective theory on the leaves in the limit $\xi \to \infty$ when dS$_4$ approaches Minkowski space. In Figure \ref{Penrose-IIa}, those leaves are represented by almost straight, essentially vertical lines connecting $B_1$ and $B_2$. They get infinitely close to the (perfectly straight) vertical line $B_1B_2$ at $\psi = \pi/2$, the timelike boundary of AdS$_5^\text{emb}$ at spatial infinity.

\bigskip

\noindent \textbf{(3) The possibility of a non-standard AdS/CFT correspondence.}  Thus, taking the above remarks together we are led to the picture of a 5-dimensional geometry with certain QFT's attached to it in a specific way. It is strikingly similar to the AdS/CFT correspondence proposed in the literature\cite{Maldacena, Gubser, Witten}. In the string theory-related correspondence, too, a theory involving gravity, namely full-fledged string theory or a low energy approximation thereof, lives on the bulk of AdS$_5^\text{}$ and is 	``holographically'' related to a CFT on the boundary, whereby the isometry group of the 5D manifold, $SO(4,2)$, acts as the conformal group on the 4D boundary \cite{Giddings, Verlinde, Heemskerk, Sathiapalan, Gao}.

The analogies to the picture based upon the functional renormalization group are  obvious. In particular the task of defining a theory of quantum gravity beyond the confines of perturbation theory is taken over by Asymptotic Safety now, replacing string theory.

The analogy  is  particularly striking if  the field theory that is defined by the asymptotically safe Type IIa trajectory $\Big\{\Gamma_k^\text{IIa}, \; k \in \mathbb{R}^+\Big\}$ is \textit{conformal}, i.e., if $\lim_{k \to 0} \Gamma_k \equiv \Gamma$ defines the action of a 4-dimensional CFT. As we mentioned in the Introduction, for the time being this is unproven in $d = 4$, but has already been established in two dimensions, where also the unitarity of the CFT was shown \cite{Andi}.

In this regard it is also important to note that recently it has been demonstrated that 4D quantum gravity based upon the action  $\int \sqrt{g}\;R$, linearized about Minkowski space, is indeed conformal, rather than merely scale invariant at the IR fixed point \cite{Farnsworth:2021zgj}. The pertinent CFT, having no stress tensor, and no relevant or marginal scalar operators, is of a non-standard type which is defined at the level of the correlation functions. This result is consistent with our expectation that the endpoint of the Type IIa trajectory is indeed a CFT.\footnote{Note however that while the limit $k \to 0$ renders Minkowski space a solution of the effective field equation, as such it does not \textit{linearize} the equation: Near the GFP, the dimensionless Newton constant scales like $g(k) \approx G_0 k^2$ with a nonzero renormalized Newton constant $G_0 \neq 0$ in general.}

\bigskip

\noindent \textbf{(4) Meaning of holography in the GEAA approach.} Within the framework of the gravitational Effective Average Action there exists a natural and perfectly general notion of \textit{holography}. It is exemplified by the above AdS/CFT picture, but its scope is much broader.

Loosely speaking, the corresponding    ``holographic principle'' conjectures, purely at a 4D level, that all actions $\Gamma_k$ at $k > 0$, including the bare one, $S \sim \Gamma_{k \to \infty}$, can be reconstructed from the standard effective action $\Gamma_{k = 0} = \Gamma$.

This is equivalent to saying that the functional RG equation defines a meaningful initial value problem also when the direction of the $k$-evolution is changed from ``downward'' to ``upward'', and the initial condition $\Gamma_{k = 0} \stackrel{!}{=}\Gamma$ is imposed in the IR rather than UV.

A solution to this ``inverse quantization problem'' extends the dynamics from the ``holographic screen'', aka, the foliation's leave at $k = 0$, into the ``bulk'' comprised of the leaves with $k >0$.

In general it may be problematic to give a mathematical meaning to such a functional variant of a boundary value problem. Along asymptotically safe RG trajectories, followed in the reversed direction, this should be possible though. A first proof of principle has appeared in ref.\cite{Manrique:2008zw} already.

\bigskip

\noindent \textbf{(5) Nonzero IR cosmological constant.} In the Type IIIa case, having $\Lambda_0 \neq 0$, the situation is different in that dS$_4$ continues to be a solution of the effective Einstein equation even when $k$ is \textit{strictly} zero. Contrary to the IIa case discussed above, the limit $k \to 0$ involves no change from the de Sitter solution  to Minkowski space at $k = 0$ at the final point of the trajectory.

Comparing Figures \ref{Penrose-IIIa} and \ref{Penrose-IIa} shows that a non-zero value  $\Lambda_0 \neq 0$ prevents the boundary of  AdS$_5^\text{emb}$ to get close to the timelike boundary of the full  AdS$_5^\text{}$. It always remains crescent-shaped when $\Lambda_0 \neq 0$, and this spoils the analogy to the AdS/CFT picture.

\subsection{The dS embedding}\label{sec:5.3}
A  geometrization of the RG flow by means of our second candidate, $\mathscr{M}_{d+1}=\text{dS}_{d+1}$, would rely upon the (dimensionless) de Sitter metric
\begin{equation}
	\di s_{d+1}^2 = (\di  \xi)^2 + \sin^2( \xi)\; \bigg[-\di t^2 + \cosh^2 (t)\; \di \Omega^2_{d-1}\,\bigg]\;.\label{I3-1}
\end{equation}
As before, let us begin by investigating the spacetime furnished with \eqref{I3-1} for the full range of coordinate values for which the metric is non-degenerate:
\begin{equation}
	\xi \;\in\; \big(0,\; \pi\big)\qquad \text{ and }\qquad 	t\;\in\; \big(-\infty,\; +\infty\big) \;.
\end{equation}
So, for a moment, we ignore the constraints due to the matching of $\xi$ with $k$.
\bigskip

\noindent \textbf{(1) Relationship $\xi$-$k$ disregarded.} To begin with we trade $\xi$ and $t$ for new coordinates, $\psi$ and $\tau$, by means of a transformation
\begin{eqnarray}
	(0,\,\pi)\times \mathbb{R} \quad&\to& \quad(0,\,\pi)\times \left(-\frac{\pi}{2},\, +\frac{\pi}{2}\right)\;,\nonumber\\
	(\xi,\,t) \quad&\mapsto&\quad \bigg(\psi(\xi,\,t),\; \tau(\xi,\,t)\bigg)\quad\;,\label{I3-3}
\end{eqnarray}
which is defined by the following functions:\footnote{All inverse trigonometric functions are understood to be principal values.}
\begin{subequations}
	\begin{align}
		\psi(\xi,\,t)\; &= \;\arccos \left(\frac{\cos(\xi)}{\sqrt{1+\sin^2(\xi)\sinh^2(t)}}\right)\;,\\
		\tau(\xi,\,t) \;&= \;\arctan \Big(\sin(\xi)\;\sinh(t)\Big)\;.
	\end{align}
\end{subequations}
This transformation recasts the metric \eqref{I3-1} in the manifest conformally flat form
\begin{equation}
	\di s_{d+1}^2 = \frac{1}{\cos^2(\tau)}\;\bigg[-\di \tau^2\;+\; \di \psi^2\;+\; \sin^2(\psi)\; \di \Omega^2_{d+1}\,\bigg]\;.\label{I3-5}
\end{equation}

\begin{figure}[t]
	\centering
	\includegraphics[scale=0.4]{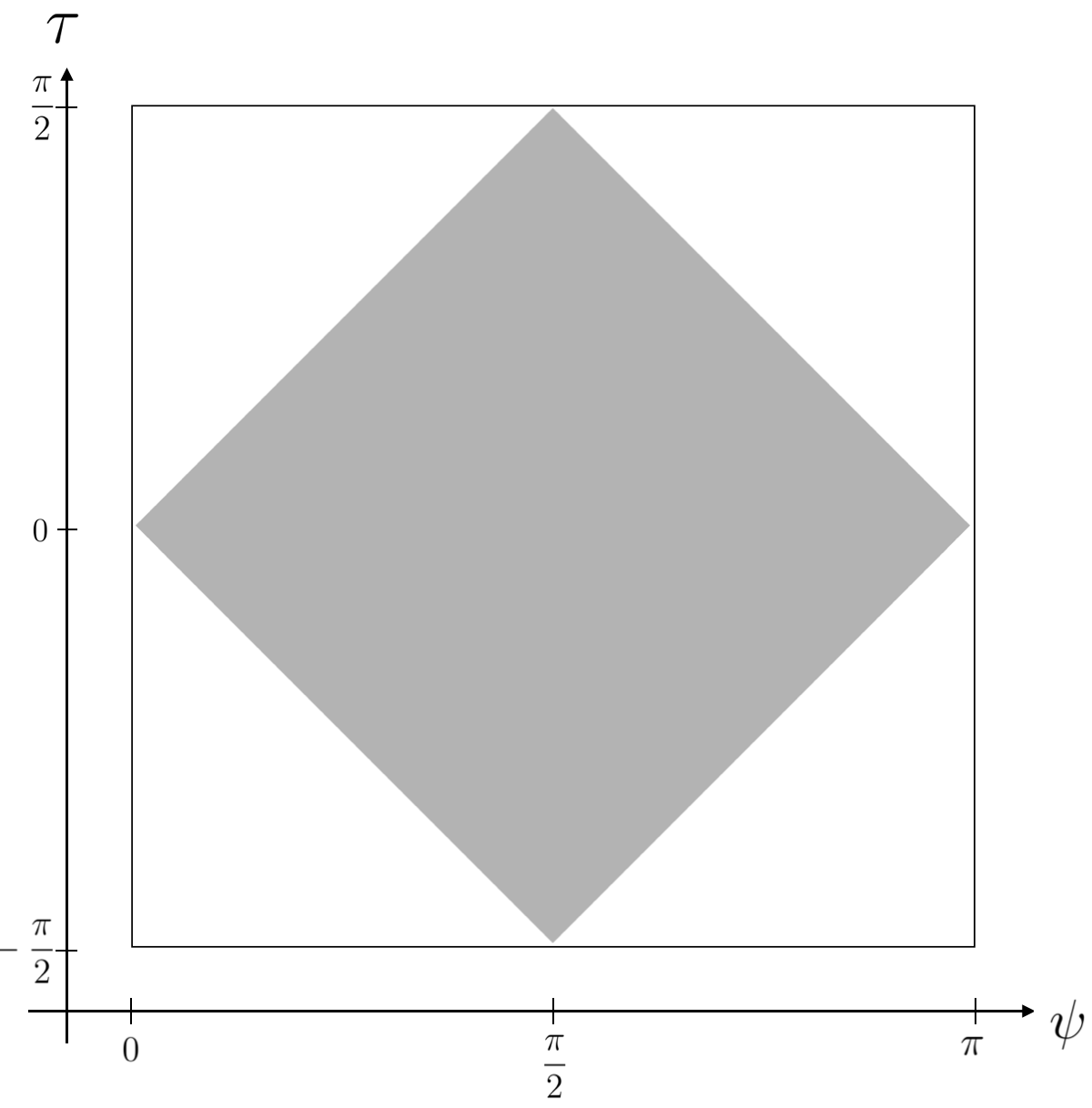}
	\captionof{figure}{The Penrose diagram of dS${}_{d+1}$ on the $\tau$-$\psi$ plane. Every point corresponds to a sphere S${}^{d-1}$ of radius $\sin^2(\psi)$. The shaded square indicates the part of the \linebreak de Sitter manifold which is covered by the $\xi$-$t$ coordinate system.}
	\label{dS-cover}
\end{figure}
\noindent Further useful properties of this coordinate transformation include
\begin{eqnarray}
	\xi \;\in\;	\left(0,\,\frac{\pi}{2}\right)\quad &\Longrightarrow&\quad \psi \;\in\;	\left(0,\,\frac{\pi}{2}\right)\;,\nonumber\\
	\xi \;\in\;	\left(\frac{\pi}{2},\,\pi\right)\quad &\Longrightarrow& \quad\psi \;\in\;	\left(\frac{\pi}{2},\,\pi\right)\;,
\end{eqnarray}
as well as $\text{sign}\big(\tau(\xi,t)\big)=\text{sign}(t)$, and
\begin{eqnarray}
	\tau(\xi, \,t = 0) = 0, \quad &\psi (\xi, \,t = 0)=\xi&, \quad \forall \;\xi \in (0,\,\pi)\;,\label{5.14}\\
	&\psi\left( \xi=\frac{\pi}{2},\, t = 0\right)=\frac{\pi}{2}&, \quad \forall \;t \in \mathbb{R}\;.\label{5.15}
\end{eqnarray}

If one allows $\psi$ and $\tau$ to freely and independently draw values from the intervals $\psi \in (0,\pi) $ and $\tau \in \left(-\frac{\pi}{2}, +\frac{\pi}{2}\right)$, respectively, \eqref{I3-5} amounts to  the familiar de Sitter metric in global coordinates, from which the Penrose diagram can be deduced in the usual way \cite{Griffiths}. However, the actual image of the above coordinate transformation is smaller than the codomain written in \eqref{I3-3}. Thus the original $\xi,t$ coordinates do not cover the de Sitter manifold fully.

\begin{figure}[t]
	\centering
	\includegraphics[scale=0.5]{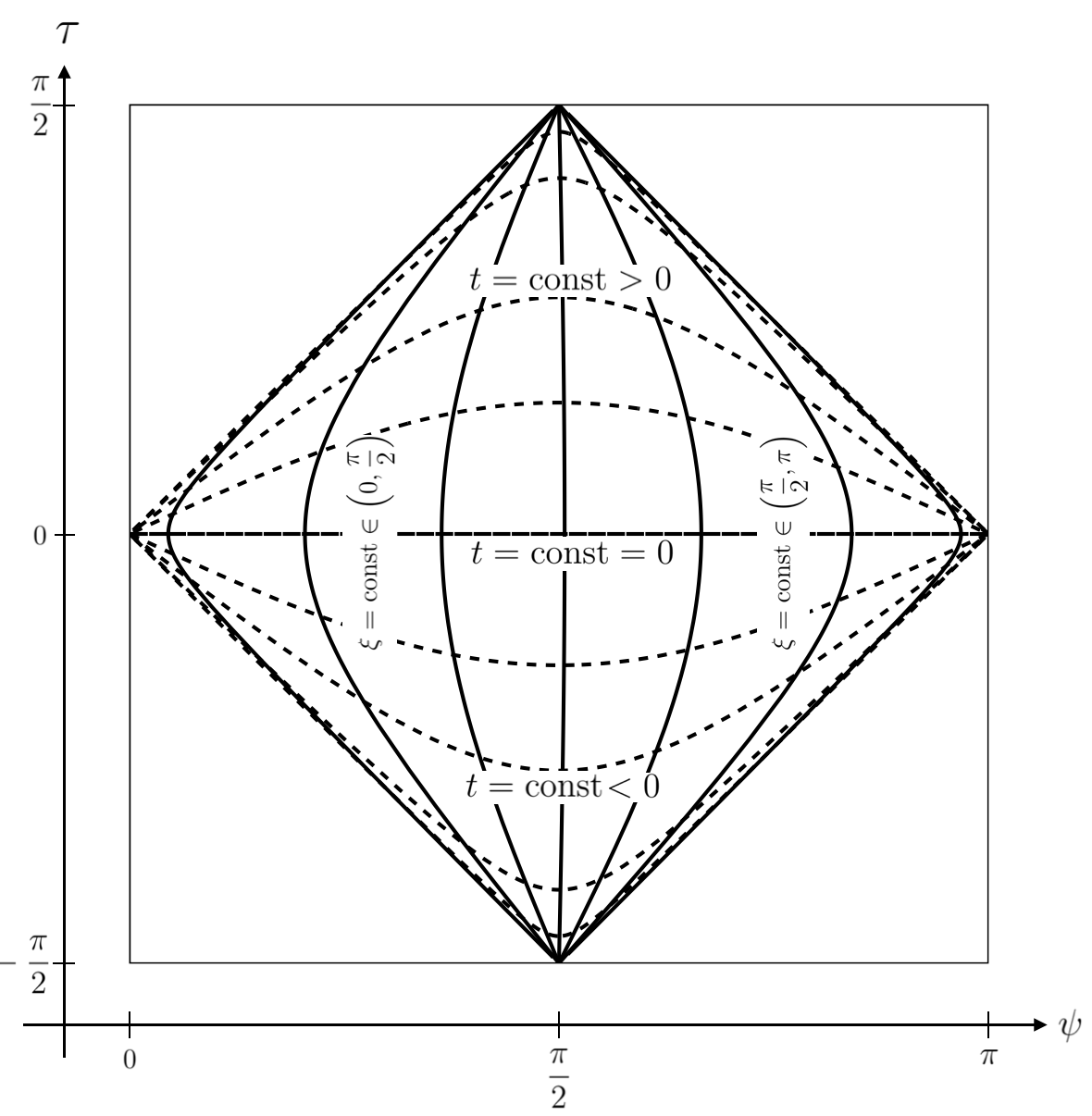}
	\captionof{figure}{The part of the dS${}_{d+1}$ Penrose diagram that is covered by $\xi$-$t$ coordinate systems. Coordinate lines with $\xi =const$ (solid lines) and $t = const$ (dashed lines) are shown.}
	\label{dS-coords}
\end{figure}

In Figure \ref{dS-cover} we draw the Penrose diagram of dS${}_{d+1}$ on the $\psi$-$\tau$ plane, and we indicate which part of spacetime is actually  covered by the $\xi,t$ coordinate system. It corresponds  to the shaded square.

Furthermore, Figure \ref{dS-coords} focuses on this particular portion of spacetime and additionally shows the net of coordinate lines with $\xi =const$ and $t =const$, respectively.

\bigskip

\noindent\textbf{(2) Imposing the $\xi$-$k$ relationship.} In Subsection \ref{sec:4.3} we concluded that the dS$_5$ embedding is viable for $y >1$ only, in which case the coordinate $\xi$ assumes values in the interval $\big(0,\, \xi_\text{max}(y)\big)$. Using \eqref{4.52} in \eqref{4.51}, we can express the upper  boundary of this interval as
\begin{eqnarray}
	\xi_\text{max}(y) \;=\; L_H^\text{R}\; \arcsin (y^{-1})\; =\; L_H^\text{R}\; \arcsin\left(\frac{L_H(0)}{L_H^\text{R}}\right)
\end{eqnarray}
Regarding its dependence on $y$, the values assumed by $\xi_\text{max}(y) $ range from $\xi_\text{max} = 0$ in the limit $y \to \infty$, to 
\begin{eqnarray}
	\lim_{y\searrow1} \;\xi_\text{max} (y)\; =\; \left(\frac{\pi}{2}\right)\;L_H^\text{R}\;,
\end{eqnarray}
when the threshold at $y = 1$ is approached.

Also in the case of the de Sitter candidate dS$_5$, the embedding of all dS$_4$-spacetimes along a complete Type IIIa trajectory covers only a  part of the 5-dimensional manifold. Henceforth denoting it by dS$^\text{emb}_5$, let us now  determine this part of the de Sitter manifold.

In Subsection \ref{sec:4.3} we saw already that, for monotonicity reasons, we must restrict the range of $\xi$, namely from the original interval $(0, \pi)$ to $\left(0, \frac{\pi}{2}\right)$, in units of $L_H^\text{R} \equiv H_\text{R}^{-1}$. Using the properties \eqref{5.14} and \eqref{5.15}, this restriction is seen to imply a corresponding restriction for the newly introduced $\psi$ coordinate, namely $\psi \in \left(0, \frac{\pi}{2}\right)$.

In geometrical terms this means that, in the Penrose diagram of Figure \ref{dS-cover}, only that subset of the shaded square is available for the geometrization which lies to the left of the vertical line at $\psi = \frac{\pi}{2}$, $\tau \in \left(-\frac{\pi}{2}, \;\frac{\pi}{2}\right)$. Thus the dS$^\text{emb}_5$ part of dS$_5$,  embedding all 4D spacetimes, must fit into a triangular region with corners at $(\tau, \psi) = (0,\,0), \;\left(\frac{\pi}{2},\,\frac{\pi}{2}\right)$, and $\left(\frac{\pi}{2},\,-\frac{\pi}{2}\right)$, respectively.

\begin{figure}[t]
	\centering
	\begin{subfigure}{0.51\textwidth}
		\includegraphics[scale=0.39]{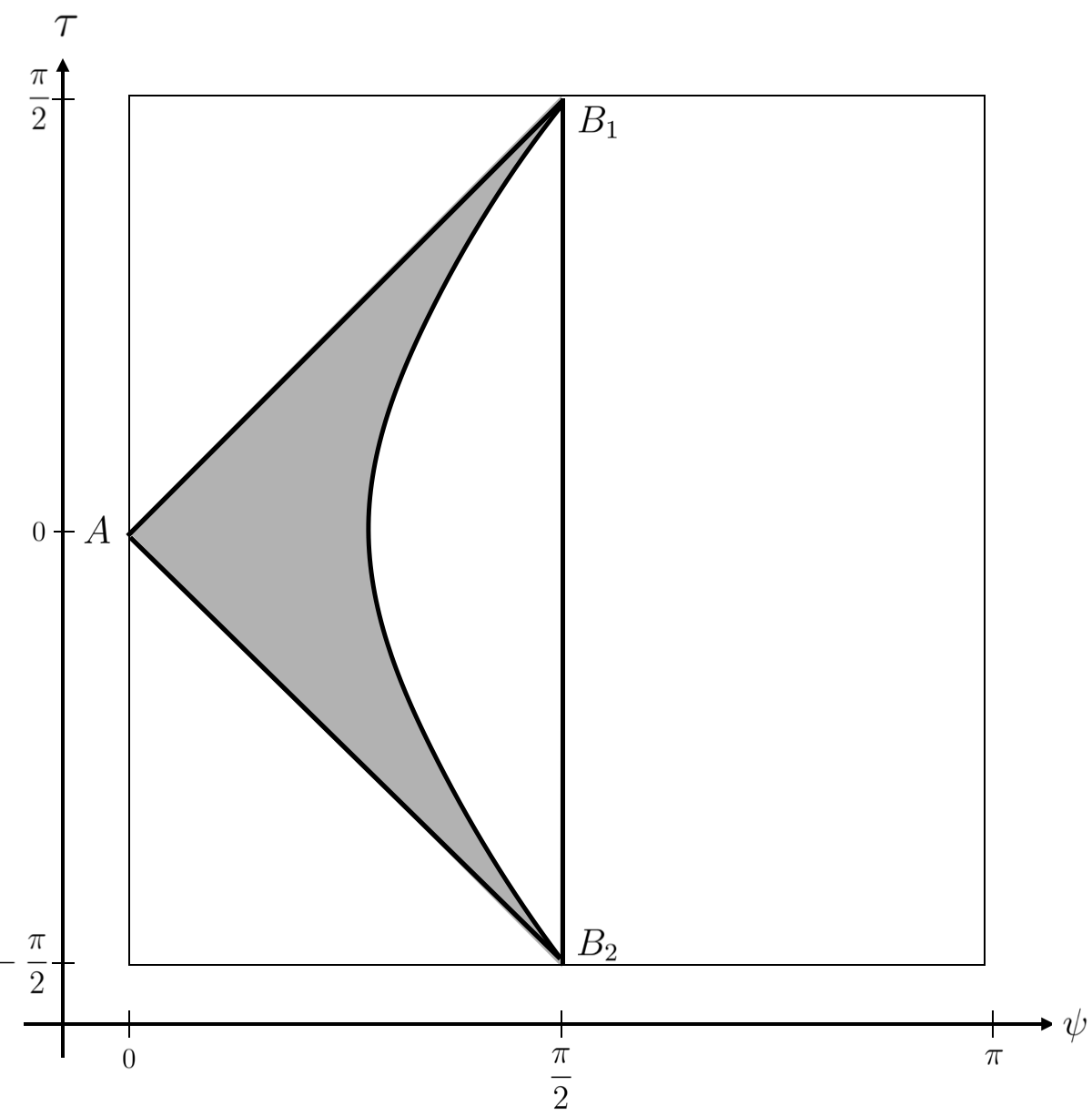}
		\caption{}		\label{dS-geometrize1}
	\end{subfigure}
	\begin{subfigure}{0.48\textwidth}
		\includegraphics[scale=0.39]{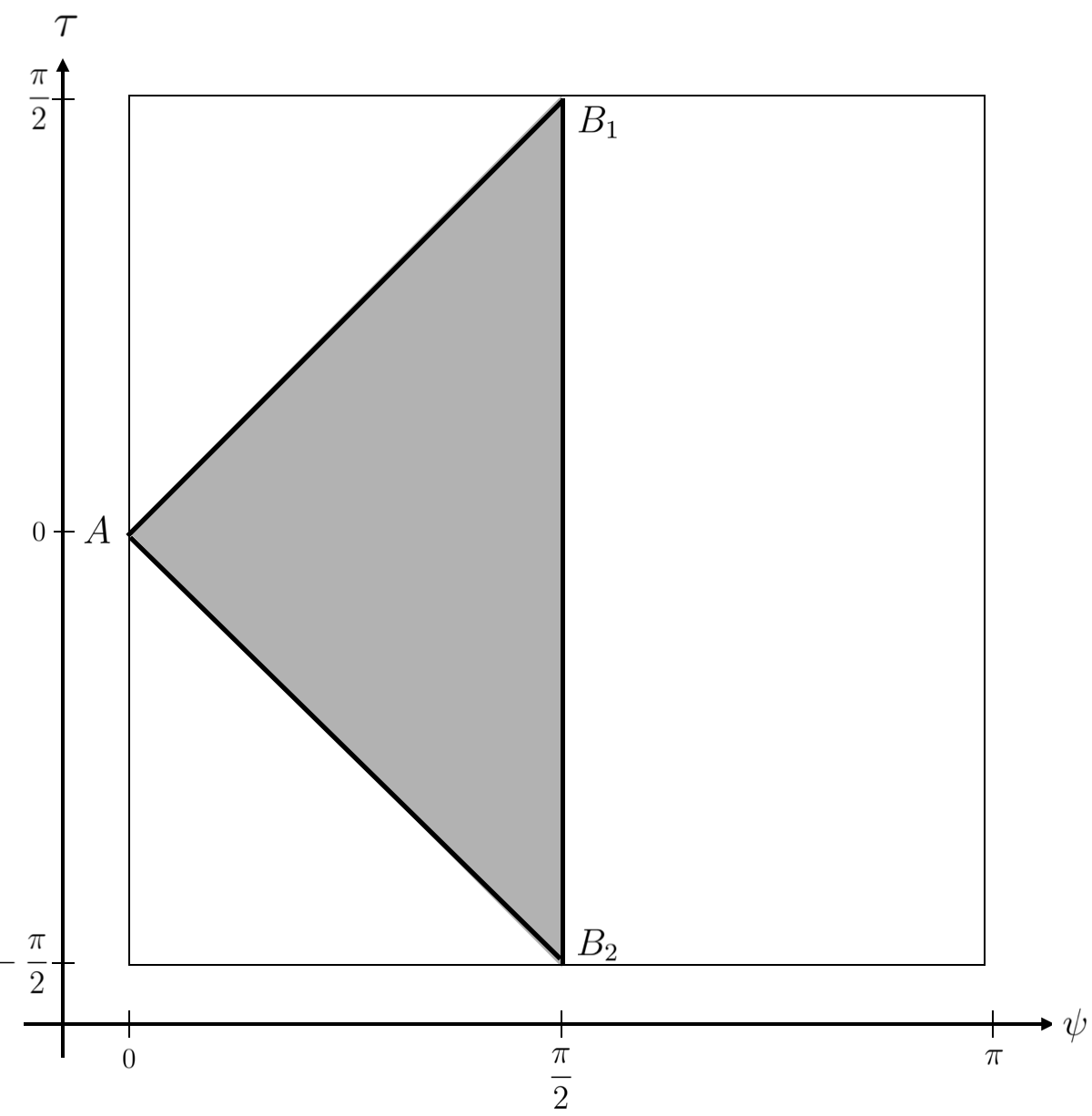}
		\caption{}		\label{dS-geometrize2}
	\end{subfigure}
	\caption{The shaded area indicates the part of the de Sitter manifold which is needed in order to fully geometrize a Type IIIa trajectory with, respectively $\Lambda_0 > \Lambda_\text{R}$ (left diagram) and $\Lambda_0 = \Lambda_\text{R}$ (right diagram).}\label{geometrize}
\end{figure}

Furthermore, taking advantage of Figure \ref{dS-coords} in order to translate the condition $\xi < \xi_\text{max}$ to the $\tau$-$\psi$ coordinate system, it becomes clear that only a subdomain of the above triangle is actually required for the embedding. In the Penrose diagram of Figure \ref{dS-geometrize1}, the subdomain is represented by the shaded crescent-shaped area. This diagram refers to a $y$-value strictly larger than unity, implying a $\bar \xi_\text{max}$ value strictly smaller than $\frac{\pi}{2}$. On the $\tau$-$\psi$ plane, this latter value is responsible for the curved boundary of the subdomain.

On the other hand, if we perform the limit $y \searrow1$, then $\bar \xi_\text{max}\nearrow\frac{\pi}{2}$, and as a consequence the curved boundary of the crescent approaches a vertical line at $\psi = \frac{\pi}{2}$. In this limit, the full triangular region is needed for the embedding. This case is depicted in Figure \ref{dS-geometrize2}.

\bigskip
\noindent\textbf{(3) The $\bm{\Lambda_0 = 0}$ problem.} Note that despite a superficial similarity of the triangular regions in, respectively, Figure \ref{dS-geometrize2} and its anti-de Sitter analog for $\Lambda_0 = 0$, the running cosmological constant on the boundary line $B_1B_2$ does not vanish for the de Sitter embedding. It rather equals the cosmological constant of the reference spacetime, $\Lambda(k)\Big|_{k = 0} \equiv \Lambda_\text{R} >0$, so that  $g_{\mu \nu}^{k = 0} = g_{\mu \nu}^\text{R}$ has nonzero curvature. The consequence is that, strictly speaking, the embedding fails for the Type IIa trajectory: The limit $\Lambda_0 \to 0$ cannot be taken \textit{while} $\Lambda_\text{R}$ \textit{is held fixed}.

\begin{figure}[t]
	\centering
	\includegraphics[scale=0.5]{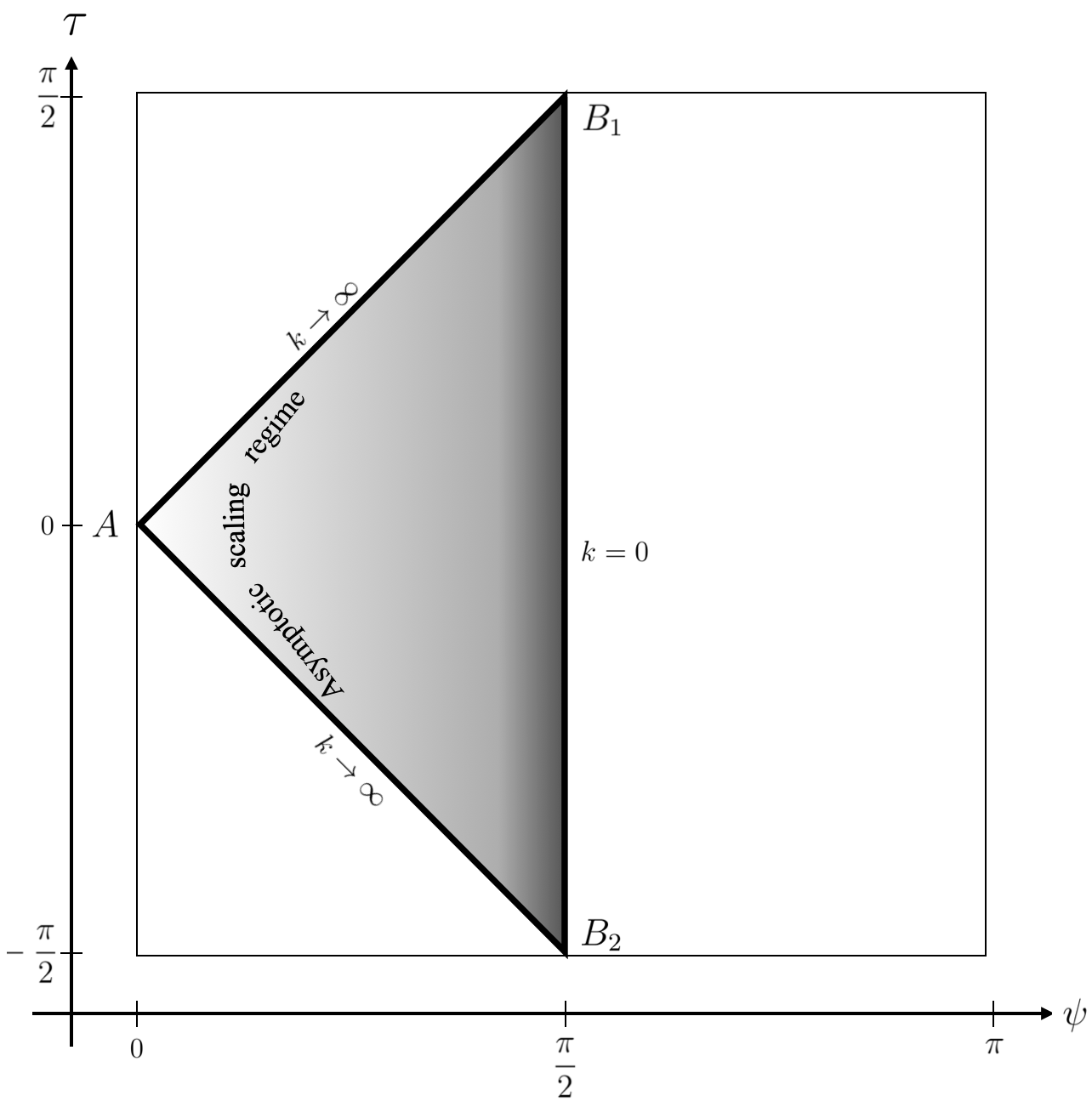}
	\captionof{figure}{Penrose diagram highlighting the triangular portion of the dS$_5$ manifold that embeds all dS$_4$ spacetimes along the Type IIa trajectories in the limit \mbox{$y= (\Lambda_0/\Lambda_\text{R})^{1/2} \searrow1$}. The gray shading indicates the local value of the RG scale $k(x^I)$.}
	\label{dS-summary}
\end{figure}

However, since $\Lambda_\text{R}$ can be given a value as small as we like, the embedding can at least approximate the geometrization of the complete Type IIa trajectory at any desired level of accuracy.

\bigskip
\noindent\textbf{(4) Interpretation and dS/CFT connection.} The status of the dS$_{5}$-candidate is less obvious than it has been for its anti-de Sitter counterpart. This concerns in particular  its role in a possible dS/CFT correspondence.

First of all, there is the following crucial difference. As illustrated in Figure \ref{dS-summary}, the boundary components of dS$_5^\text{emb}$ include a 4D timelike component with RG parameter $k = 0$, as did those of AdS$_5^\text{emb}$. Now, while in the anti-de Sitter case this component was a boundary of both AdS$_5^\text{emb}$ and the complete AdS$_5^\text{}$ manifold, this is not so in the de Sitter case: The Penrose diagram of Figure \ref{dS-summary} represents the component in question  by the (open) line segment $B_1B_2$, and obviously it entirely consists of \textit{inner} points of dS$_5^\text{}$ only.

As a consequence, there seems to be no natural way of linking the fully quantized asymptotically safe theory, governed by $\Gamma_{k \to 0}$, to the  boundary at spatial infinity of  de Sitter space proper.

This fact motivates invoking the early and late time boundaries of dS$_5^\text{}$ instead. Indeed, in Figure \ref{dS-summary}, the two horizontal lines represent the spacelike past and future  infinity of dS$_5^\text{}$, respectively. The triangle representing dS$_5^\text{emb}$ touches them in the points $B_1$ and $B_2$, which amount to 3-spheres actually.

The remarkable situation in the 3D spaces at $B_1$ and $B_2$ is appreciated best in Figure \ref{dS-coords}, where the $t$- and $\xi$-coordinate lines are shown on the $\tau$-$\psi$ plane. On the one hand, the S$^3$'s at $B_1$ and $B_2$ are seen to correspond to the early and late time limits $t \to \pm \infty$. But on the other hand, also \textit{all} the $\xi = const$ lines accumulate at $B_1$ and $B_2$, i.e., the leaves of constant scale. This includes even the limiting leave of the foliation, $\xi = const \to 0$, which is ruled by the NGFP action $\Gamma_{k \to \infty}$.

From the effective field theory perspective, it is quite remarkable that  the actions $\Gamma_k$ with $k = k(\xi)$ and $t \to \pm \infty$, $\xi \in \left(0, \,\frac{\pi}{2}\right)$ all seem to ``meet'' at $B_1$ and $B_2$. It is natural to interpret this situation by saying that all effective actions $\Gamma_k$, $k \in \mathbb{R}^+$, are  equally relevant there, and that therefore the geometry and matter fluctuations of all scales must be equally important. Clearly this is nothing but the standard characterization of a critical phenomenon at some RG fixed point.

Thus, if we hypothesize again that the QEG fixed points are conformal, the overall conclusion is that the S$^3$ spaces at $t \to \pm \infty$ indeed carry a 3D conformal field theory related to the GEAA in its early/late time regime.

Remarkably, this  type of a ``dS/CFT correspondence'' which emerges here as a special solution to the QEG flow and field equations has  essentially the same general structure as the one proposed in the literature on the basis of entirely different arguments \cite{Strominger}.

\section{The standard AdS/CFT correspondence:\\a comparison}\label{sec:new6}
Above we saw that the result of the proposed geometrization procedure has a number of features in common with the usual AdS/CFT correspondence based upon string theory.
Nevertheless, there are also marked differences between the string theory- and the GEAA-based picture, respectively. In this subsection we are going to compare the two approaches in some detail, highlighting their similarities and dissimilarities. We focus on the anti-de Sitter case here; for the dS/CFT correspondence the situation is analogous in most regards.

\bigskip
\noindent\textbf{(1) More than pure kinematics.} It is a well-known fact, predating the AdS/CFT correspondence, that the conformal group of a CFT in $d$-dimensions is realized by isometries of a $(d+1)$-dimensional hyperbolic space. Therefore any construction that starts out from a CFT, adds an extra dimension, and furnishes the higher dimensional spacetime with (standard) gravity while preserving the symmetries, is bound to find an anti-de Sitter space \cite{Brown}.

It is important to emphasize that the GEAA approach goes beyond this simple kinematic realization of the symmetries present, for the following reasons:

\begin{enumerate}[label=\textbf{(\roman*)}]
\item Our main result, the fact that the RG trajectory considered geometrizes by means of an AdS$_5$ space, was obtained without any assumption about whether or not the scale invariance at the NGFP and the GFP extends to full conformal symmetry. It does not rely on the boundary theory being a CFT, and is equally true in the (unlikely) case that the NGFP, or the GFP, or both are not conformal. Hence the geometrization which we found reveals first of all a general AdS/QFT relationship which may or may not be a AdS/CFT one, depending on the properties of the GFP.

\item Conversely, our construction does depend on critical properties of its input from the FRG. The latter encodes information about the theory's \textit{dynamical} properties. In particular the choice of a Type IIa trajectory, and the monotonicity of its running $\Lambda(k)$ were indispensable in order to obtain AdS$_5$.

\item Unlike the standard AdS/CFT correspondence, our approach to the geometrization of RG flows does \textit{not} yield \textbf{5D} Einstein-Hilbert gravity on AdS$_5$. Rather, every 4-dimensional leave of the AdS$_5$ foliation carries its own copy of \textbf{4D} Einstein-Hilbert gravity, with parameter values $G(k)$ and $\Lambda(k)$ depending on the leave.\\
By contrast, in the usual AdS/CFT correspondence, applying the holographic RG to the energy momentum tensor perturbation in a flat boundary $O(N)$ model (without gravity!) is known to give rise to a \textit{propagating graviton in the bulk} - something that is not found here. Within the geometrization construction the \textbf{5D} Einstein equation does not play any particular role. The basic field equations rather consist of an infinite stack of \textbf{4D} Einstein equations labeled by $k$.\\
As a consequence, instead of one \textbf{5D} graviton, we deal with an infinite family of \textbf{4D} gravitons, which do not get coupled by the \textit{field equations}. Rather, it is the \textit{FRG flow equation} that connects the leaves of the foliation and their respective gravitons at different scales $k$ and $k + \text{d}k$. Since our entire geometrization is based upon \textit{one single} solution to the FRG equation, namely the selected RG trajectory chosen as the input, the relation between neighboring leaves appears to be ``non-dynamical'' from the point of view of the field equations.
\item In this context we also mention that from the GEAA viewpoint of the present paper the astonishing ``miracle'' behind the standard AdS/CFT correspondence is that, \textit{under very special conditions}, the universally applicable but enormously complicated FRG equation can be replaced with something  much simpler, namely the \textbf{5D} Einstein equation.\footnote{Clearly in most other connections one hardly would call Einstein's equation ``simple''. And yet it is true that the FRGE is a by far more complex mathematical object: It is a \textit{functional} equation, it is highly \textit{non-local}, and it is an \textit{integro-differential} equation.} One of the motivations for the present work is the hope that ultimately one might be able to actually \textit{derive} what these specific conditions are, and thus understand better why the usual AdS/CFT correspondence works for some gravity-matter systems but not for others.
\end{enumerate}

\bigskip

\noindent\textbf{(2) Duality.} A key property of the standard AdS/CFT correspondence is that it constitutes a duality, in the sense that the boundary values of the gravity etc. fields that live in the bulk act as sources of the boundary operators. The duality provides a map between two dynamical theories that are described by two actions that look very different and involve different degrees of freedom. One of them includes dynamical gravity, the other does not. The map leads to a one-to-one relationship between gravity fields and operators of a flat space QFT on the boundary \cite{Haro}.

Within the GEAA-based geometrization approach, no duality of this sort has emerged so far, at least not within the specific example worked out in the present paper.

The situation may change however once in the future more complex RG trajectories are considered which include a nontrivial matter part added to the running gravitational action. At least in principle, something comparable to a duality in the sense above is given a chance then to arise dynamically, provided one employs the generalized FRG approach, already well developed for flat space, which allows for a continuous change of the dynamical variables during the RG evolution. A typical example of this kind is in QCD the transition from quarks and gluons to hadrons \cite{EWG}.

Whether or not this extended framework is employed, there always remains a main difference between standard AdS/CFT and the GEAA geometrization approach: In the former case the boundary theory is a pure matter theory on flat space, while in the latter it includes dynamical gravity, \textit{quantum} gravity even.

In fact, the present approach deals with dynamical, quantized gravity in the boundary, and this goes beyond what is done in the usual AdS/CFT correspondence.  For technical simplicity the scale dependent actions considered here are even \textit{only} for gravity, containing no matter fields for the time being (see below).
\bigskip

\noindent\textbf{(3) Holographic RG.} The idea of combining the AdS/CFT setting with the RG, interpreting the additional dimension as the RG scale, is an old one \cite{Akhmedov, Verlinde, Bianchi}. In the usual AdS/CFT correspondence, one considers an UV-CFT ``living'' on the boundary and one perturbs this fixed point field theory by relevant operators, thus generating an RG flow towards the IR.

In the GEAA geometrization approach the situation is different. In the first step we derived an ``AdS/QFT correspondence'' which involves
\begin{enumerate}[label=\textbf{(\roman*)}]
\item a UV-QFT given by the NGFP, i.e., the fixed point related to the nonperturbative renormalizability;
\item an IR-QFT which is defined by the GFP and lives on the boundary.
\end{enumerate} 
Then we argued that both the UV-QFT and the IR-QFT are likely to be CFT's, but this has not been proven yet. Clearly the similarity with the traditional string theory related setting is strongest if at least the IR-QFT is conformal; it would then define the ``CFT'' appearing in the designation ``AdS/CFT''.

By contrast, our UV-QFT as given by the non-Gaussian fixed point has no counterpart in the traditional AdS/CFT correspondence. The existence of the NGFP is the very hallmark of Asymptotic Safety which, rather than string theory, provides the UV completion in the present case.

\bigskip
\noindent\textbf{(4) Internal space.} Besides the UV completion, string theory plays yet another role in the usual AdS/CFT scenario: In addition to AdS$_5$, the string theory construction in 10 dimensions supplies an additional internal space, for example S$^5$ in the simplest case. The internal space fixes further symmetries beyond conformal symmetry and thus determines the field content of the theory involved.

On the GEAA side, a similar role is played by the choice of the \textit{theory space} the functional RG equation is operating upon. For different such choices, the pertinent functionals $\Gamma_k\left[g_{\mu \nu}, \textit{matter fields}, \cdots\right]$ depend on different sets of matter fields, and therefore entail RG flows that are to be computed from different functional RG equations. Hence, the component form of the functional RG trajectories, $k \mapsto\left(G(k), \Lambda(k), \cdots, u_1(k), u_2(k), \cdots\right)$, will include additional running couplings and masses, $u_i(k)$. They reflect the invariants that can be built from the specific set of matter fields chosen.

In the present paper we analyzed the geometrization in the simplest case only, namely for pure gravity, since its RG flow is fairly well understood by now \cite{Frank}. However, it will be interesting to explore also possible geometrizations of matter-coupled gravity in future work.
The above AdS$_5\times$S$^5$ example, for instance, determines the boundary CFT to be $ N=4$ Super-Yang-Mills theory. It is highly intriguing that this theory may emerge from the GEAA approach as a $k \to 0$ limit. For the time being this system is still beyond our computational possibilities, but steady progresses is made in this direction \cite{10people}.

\section{Summary and conclusion}\label{sec:6}
When applied to quantum systems that are able to predict the geometry of the spacetime $\mathscr{M}_d$ they live in, the scale dependent actions $\Gamma_k$ provided by the functional renormalization group imply a set of coupled effective field equations which govern the expectation values of the  gravitational and matter fields. Their  scale dependent solutions include an effective metric tensor $g_{\mu \nu}^k$. Heuristically, it can be regarded as the description of a coarse-grained, fractal-like spacetime on a variable resolution scale determined by the value of the RG parameter $k$.

In this paper we advocated a geometrization of the resulting family of pseudo-\linebreak Riemannian structures $\Big(\mathscr{M}_d,\, g_{\mu \nu}^k\Big)$, $k \in \mathbb{R}^+$. It consists in isometrically embedding them into a  higher-dimensional manifold $\Big(\mathscr{M}_{d+1},\, {}^{(d+1)}g_{IJ}\Big)$, which then encapsulates the entire information about the RG flow in one single ``scale-space-time''. In a coordinate independent fashion, it can be thought of as a foliated manifold whereby the leaves of the foliation describe the ordinary spacetime on varying resolution scales.

Typically different generalized RG trajectories (solutions to the combined  RG + Einstein equations) will lead to different ($d+1$)-dimensional manifolds. It is therefore an intriguing question which types of such embedding manifolds can actually occur in a given fundamental theory of quantum gravity.

We addressed this question within 4D Quantum Einstein Gravity which (almost certainly) is asymptotically safe. It owes its nonperturbative renormalizability to a non-Gaussian fixed point, and its RG flow also features a second, Gaussian fixed point. While the fixed point theories are scale invariant, full conformal invariance is likely, but has not been demonstrated so far. 

In the present paper we extended the work initiated in [I] by generalizing the flat and Ricci flat embedding manifolds studied there to  arbitrary ($d+1$)-dimensional Einstein spaces, and by allowing for a Lorentzian signature of the to-be-embedded spacetimes. Then, imposing maximum symmetry on them, we demonstrated that from Asymptotic Safety there arise  two solutions for the embedding space, namely certain parts of the AdS$_5$  and the dS$_5$ manifold,  which we denoted AdS$_5^{\text{emb}}$ and dS$_5^{\text{emb}}$, respectively.

The 5-dimensional picture that emerges is particularly striking when combined with the plausible assumption that the QEG fixed points are indeed conformal. Then the foliation furnishing (A)dS$_5^{\text{emb}}$ establishes a relationship between quantum theories in the bulk and on the boundary of (A)dS$_5$ which has exactly the same structure as in the well-known (A)dS/CFT correspondences that have been extensively discussed in the literature.

Both the (A)dS/CFT and the Asymptotic Safety approach rely upon a specific UV complete gravity theory in the bulk. In our case this role is played by QEG, nonperturbatively renormalized by means of the Type IIa trajectory, i.e., the one that crosses over from the NGFP to the GFP. Its endpoint has a vanishing cosmological constant, $\lim_{k\searrow0}\Lambda(k) = 0$, and so the corresponding solution to Einstein's equation is (3+1)-dimensional Minkowski space. Hence the effective action which has all fluctuation modes integrated out would amount to a CFT which lives on flat space.

In the anti-de Sitter case, the habitat which the foliation allocates to this CFT is the timelike 4D spatial boundary of AdS$_5$. For the de Sitter embedding, it consists instead of the S$^3$ spaces (without time) located at the spacelike past and future infinity of dS$_5$. The properties of the emergent bulk/boundary connections have been detailed and discussed in Sections \ref{sec:5.2} and \ref{sec:5.3} already.

In summary, the findings in this paper strongly support the idea that, at least in principle, it may be possible to discover various forms of 	``(A)dS/CFT correspondences'' as specific solutions to the flow and field equations of matter-coupled QEG. While this may include the known examples\footnote{At the level of a field theory approximation to string theory.}, the present approach has the potential of identifying  new ones also. Clearly in practice such an endeavor is beset  
with a large number of technical difficulties. In particular much more general truncations in theory space must be used which also should be able to discriminate between different matter systems. But nevertheless, ultimately this approach may help in better understanding the \textit{raison d'être} of the known bulk/boundary correspondences.

\subsection*{Acknowledgments}
We are grateful to Roberto Percacci for inspiring discussions and for bringing ref.  \cite{Farnsworth:2021zgj} to our attention. We also would like to thank Bianca Dittrich and Alessia Platania for helpful comments on the manuscript.
RF gratefully acknowledges the hospitality of Perimeter Institute on her visit during which part of this work was completed. This research was supported in part by Perimeter Institute for Theoretical Physics. Research at Perimeter Institute is supported by the Government of Canada through the Department of Innovation, Science and Economic Development and by the Province of Ontario through the Ministry of Research and Innovation.

\newpage
\begin{appendices}
	\section{The GEAA approach to \\quantum gravity}
	The gravitational effective average action (GEAA) used in this paper is a functional RG approach to quantum gravity which has mostly been used in the context of Asymptotic Safety, whose applicability is much broader though \cite{Martin, Frank}. The framework is diffeomorphism invariant and, as is mandatory in fundamental theories of gravity, it is fully Background Independent at the physical level.
	
	When applied to an arbitrary set of matter fields coupled to (metric) gravity, it is formulated in terms of scale dependent effective actions\footnote{The dots stand for ghosts, auxiliary fields and BRST sources which are not essential here \cite{Martin}.} $\Gamma_k \left[g_{\mu \nu}, \psi, \cdots; \bar g_{\mu \nu}\right]$ which, besides the usual  expectation values $g_{\mu \nu} \equiv \langle \hat g_{\mu \nu}\rangle, \cdots$, depend on an additional argument, namely an \textit{arbitrary} background metric. Background Independence is implemented, at each scale $k$ independently, by fixing $\bar g_{\mu \nu}$ \textit{dynamically}, rather than by fiat.
	
	The condition for a self-consistent, that is, dynamically selected background metric $\left(g^\text{sc}_k\right)_{\mu \nu}$ is the tadpole condition for the fluctuation $\hat h_{\mu \nu} \equiv \hat g_{\mu \nu} - \bar g_{\mu \nu}$, i.e., $\langle\hat h_{\mu \nu}\rangle_{\bar g_k^\text{sc}} =0\;\Leftrightarrow\; \langle\hat g_{\mu \nu}\rangle_{\bar g_k^\text{sc}} =\left(\bar g_{\mu \nu}^\text{sc}\right)_{\mu \nu}$, or, explicitly,
	\begin{equation}
	\left.	\frac{\delta}{\delta h_{\mu \nu}(x)} \Gamma_k \left[\bar g_{\mu \nu}+h_{\mu \nu}, \psi,\cdots; \bar g_{\mu \nu}\right]\right|_{h=0, \;\bar g = \bar g^\text{sc}_k} = 0\,.\label{sc}
	\end{equation}
	In the GEAA approach, the condition \eqref{sc}, along with analogous equations for the other dynamical fields, constitutes the basic effective field equations which generalize the classical Einstein and matter field equations. Within simple approximations (``single metric truncations'') such as the Einstein-Hilbert ansatz \eqref{trunc}, they boil down to ordinary equations of motion of the Euler-Lagrange type which involve only one metric variable.
	
	The above expectation values $\langle \cdots\rangle_{\bar g}$ possess a dependence on both the RG scale $k$, an IR cutoff, and the background metric $\bar g_{\mu\nu}(x)$, which they inherit from the underlying functional integral.
	
	The action functional $\Gamma_k$ is formally defined by a modified BRST-gauge fixed functional integral over the c-number analog $\hat g_{\mu \nu}$ of the metric operator. It is written as an integration over an appropriate fluctuation variable $\hat h_{\mu \nu}$ that parameterizes the deviation of $\hat g_{\mu \nu}$ from $\bar g_{\mu \nu}$. There are many ways of introducing $\hat h_{\mu \nu}$; only in the simplest case of a linear background splits one sets $\hat h_{\mu \nu} \equiv \hat g_{\mu \nu} - \bar g_{\mu \nu}$. Given a diffeomorphism invariant bare action $S\left[\hat g, \cdots\right]$, one adds gauge fixing and ghost terms, introducing Faddeev-Popov fields $C^\mu$ and $\bar C_\mu$, and then considers a generating functional of the form
		\begin{equation}
	W_k\left[J; \bar g\right] = \log \bigintssss \mathscr{D}\hat\varphi \exp \left\{-S_\text{tot}\left[\hat \varphi; \bar g\right]+\int \text{d}^dx \sqrt{\bar g} J_i\hat \varphi^i\right\} \exp\bigg\{-\Delta S_k\left[\hat \varphi; \bar g\right] \bigg\}\label{genfun}
	\end{equation}
	Here $\hat \varphi \equiv \left(\hat\varphi^i\right)\equiv \left(\hat h_{\mu \nu}, \hat \psi, C^\mu, \bar C_\mu, \cdots\right)$ denotes the set of all dynamical fields, and $J = \left(J_i\right)$ are sources coupled to them. The total action $S_\text{tot}$ includes $S\big[\bar g + \hat h, \cdots\big]$ plus the gauge fixing and ghost contributions.
	
	The non-standard piece in \eqref{genfun} is the cutoff action $\displaystyle\Delta S_k \equiv \frac{1}{2}\int \text{d}^d x\sqrt{\bar g}\hat \varphi(x)\mathcal{R}_k \hat \varphi(x)$ which serves the purpose of installing an IR cutoff in the functional integral by  giving a mass $\propto k$ to all those modes of $\hat \varphi$ that possess a (\textit{covariant momentum})$^2$ smaller than $k^2$. The essential ingredient in the pseudo-differential operator $\mathcal{R}_k\left[\bar g\right] \propto k^2 R^{(0)}\left(-\Box_{\bar g}/k^2\right)$ constructed from the Laplacian on the background geometry, $\Box_{\bar g} \equiv \bar g^{\mu \nu}\bar D_{\mu} \bar D_{\nu}$. Its eigenvalue determines the (\textit{covariant momentum})$^2$ of the respective eigenmode. Furthermore, $R^{(0)} (\cdot)$ is a largely arbitrary interpolation function. It is required to decrease monotonically from $R^{(0)}(0)=1$ to $R^{(0)}(\infty)=0$, displaying a smooth transition from the suppressed (``massive'') to the un-suppressed (``massless'') regime near $-\Box_{\bar g} /k^2 \approx 1$.
	
	The action functional $\Gamma_k \left[\varphi; \bar g\right]$ is defined as the Legendre-Fenchel transform of $W_k \left[J;\bar g\right]$ with respect to $J$, at fixed $k$ and $\bar g_{\mu \nu}$, with $\Delta S_k\left[\varphi;\bar g\right]$ subtracted from it.
	
	On the basis of the regularized functional integral a number of properties of $\Gamma_k$ can be proven: $\Gamma_k$ is invariant under diffeomorphisms acting on both $\varphi$ and $\bar g_{\mu \nu}$, it satisfies modified BRST- and split-symmetry Ward identities, and it obeys an exact functional integro-differential equation at fixed $k$. The latter entails that (roughly speaking) the GEAA interpolates in theory space between the classical action $\underset{k \to \infty}{\lim} \Gamma_k = S_\text{tot}$ and the ordinary effective action $\underset{k \to 0}{\lim}\Gamma_k = \Gamma$. Most importantly, the GEAA satisfies an exact functional RG equation which governs its $k$-dependence:
	\begin{equation}
	\partial_k \Gamma_k \left[\varphi ; \bar g\right] = \frac{1}{2}\text{STr} \left[\left(\Gamma^{(2)}_k\left[\varphi; \bar g\right] + \mathcal{R}_k\left[\bar g\right]\right)^{-1} \partial_k \mathcal{R}_k \left[\bar g\right]\right]\label{flow}
	\end{equation}
	This flow equation has a similar formal structure as the Wetterich equation for matter fields \cite{Wetterich-0, Wetterich+1, Wetterich+2}. However, in eq.\eqref{flow}, $\Gamma^{(2)}_k$ stands for the Hessian of $\Gamma_k$ with respect to the dynamical fields $\varphi^i = \left(h_{\mu \nu}, \cdots\right)$ only, all derivatives being taken fixed $\bar g_{\mu \nu}$. This difference in the status of the arguments $\varphi$ and $\bar g_{\mu \nu}$, respectively, reflects the unavoidable ``bi-metric'' character of the GEAA: It depends on $\bar g_{\mu \nu}$ and $g_{\mu \nu}\equiv \bar g_{\mu \nu}+h_{\mu \nu}$ as two in principle completely independent metrics.
	
	 In practice, this property causes considerable computational challenges. They are the very price we must pay for Background Independence if, at the same time, we want to take advantage of actually having a background geometry available as a tool during the quantization process.
	
	For further details the reader is referred ref.s \cite{Martin, Frank}. Therein also the application of the flow equation \eqref{flow} to the example of the Einstein-Hilbert  approximation is described in explicit detail. For a first confirmation of its reliability by means of a truncation ansatz admitting a nontrivial dependence on two independent metrics, see refs. \cite{Manrique, Daniel}.
	
\section{An approximation for $\bf{\Lambda(k)}$}\label{app}

The relevant features of the \textit{dimensionful} cosmological constant $\Lambda(k)$, along Type IIa and IIIa trajectories in $d = 4$, are well described by the following analytic approximation \cite{Martin, Frank1}:
	\setcounter{secnumdepth}{1}
\begin{empheq}[
	left=
	{	\Lambda(k) \;\approx\;  } 
	\empheqlbrace
	]{align}
	\displaystyle\Lambda_0 + \nu \;G_0\;k^4\qquad \text{for } \;\;\qquad 0\leq k\lesssim\hat k\label{ap1}
	\\\;\;\;\;\lambda_\ast \; k^2\qquad\qquad\text{ for }\quad \;\;\;\; \hat k \lesssim k <\infty \label{ap2}
\end{empheq}
The constants $\nu$ and $\lambda_\ast$ are an output of the RG equations, while $G_0 = G(k = 0)$ and $\Lambda_0 = \Lambda(k = 0)$ are constants of integration selecting the RG trajectory. The scale
\begin{equation}
	\hat k \;=\; \left(\frac{\lambda_\ast}{\nu\; G_0}\right)
\end{equation}
marks the transition from the semiclassical regime of the trajectory to the asymptotic scaling regime of the non-Gaussian fixed point.

Introducing an arbitrary reference scale $k_\text{R}$ at which $\Lambda(k_\text{R}) \equiv \Lambda_\text{R}$, the approximation yields the following ratio of cosmological constants:
\begin{empheq}[
	left=
	{	Y(k)\;=\; \ffrac{\Lambda(k)}{\Lambda_\text{R}} \;\approx\;  } 
	\empheqlbrace
	]{align}
	\;y^2 + \;\big(\ell\;k\big)^4\qquad\; \text{for } \;\;\qquad 0\leq k\lesssim\hat k\label{ap3}
	\\ \;\big(L\;k\big)^2\qquad\qquad\text{ for }\quad \;\;\;\; \hat k \lesssim k <\infty \label{ap4}
\end{empheq}
Herein the length scales $\ell$ and $L$ are given by, respectively,
\begin{equation}
	\ell \; \equiv\; \left(\frac{	\nu\; G_0}{\Lambda_\text{R}}\right)^{1/4}\qquad \text{and} \qquad L \; \equiv\; \left(\frac{\lambda_\ast}{\Lambda_\text{R}}\right)^{1/2}\;.
\end{equation}
Furthermore, $y ^2\equiv Y(0)=\Lambda_0/\Lambda_\text{R}$ coincides with the dimensionless parameter introduced in the main text.
\end{appendices}

\end{document}